\documentclass[11pt]{article}
\usepackage{amssymb,amsmath,amsfonts}
\usepackage{graphicx}
\usepackage{graphics}
\usepackage{eepic,epsfig}

\@addtoreset{equation}{section}

\makeatother

\begin{document}

\title{Vacuum currents in braneworlds on AdS bulk\\
with compact dimensions}
\author{ S. Bellucci$^{1}$\thanks{%
E-mail: bellucci@lnf.infn.it }, A. A. Saharian$^{2}$\thanks{%
E-mail: saharian@ysu.am }, V. Vardanyan$^{2}$ \vspace{0.3cm} \\
\textit{$^1$ INFN, Laboratori Nazionali di Frascati,}\\
\textit{Via Enrico Fermi 40,00044 Frascati, Italy} \vspace{0.3cm}\\
\textit{$^2$ Department of Physics, Yerevan State University,}\\
\textit{1 Alex Manoogian Street, 0025 Yerevan, Armenia }}
\maketitle

\begin{abstract}
The two-point function and the vacuum expectation value (VEV) of the current
density are investigated for a massive charged scalar field with arbitrary
curvature coupling in the geometry of a brane on the background of AdS
spacetime with partial toroidal compactification. The presence of a gauge
field flux, enclosed by compact dimensions, is assumed. On the brane the
field obeys Robin boundary condition and along compact dimensions
periodicity conditions with general phases are imposed. There is a range in
the space of the values for the coefficient in the boundary condition where
the Poincar\'{e} vacuum is unstable. This range depends on the location 
of the brane and is different for the regions
between the brane and AdS boundary and between the brane and the horizon. In
models with compact dimensions the stability condition is less restrictive
than that for the AdS bulk with trivial topology. The vacuum charge density
and the components of the current along non-compact dimensions vanish. The
VEV\ of the current density along compact dimensions is a periodic function
of the gauge field flux with the period equal to the flux quantum. It is
decomposed into the boundary-free and brane-induced contributions. The
asymptotic behavior of the latter is investigated near the brane, near the
AdS boundary and near the horizon. It is shown that, in contrast to the VEVs
of the field squared and energy-momentum tensor, the current density is
finite on the brane and vanishes for the special case of Dirichlet boundary
condition. Both the boundary-free and brane-induced contributions vanish on
the AdS boundary. The brane-induced contribution vanishes on the horizon and
for points near the horizon the current is dominated by the boundary-free
part. In the near-horizon limit, the latter is connected to the
corresponding quantity for a massless field in the Minkowski bulk by a
simple conformal relation. Depending on the value of the Robin coefficient, 
the presence of the brane can either increase or decrease the vacuum currents.
Applications are given for a higher-dimensional version of the 
Randall--Sundrum 1-brane model.
\end{abstract}

\bigskip


\bigskip

\section{Introduction}

\label{sec:introd}

Anti-de Sitter (AdS) spacetime is one of the simplest and most interesting
spacetimes allowed by general relativity. It is the unique maximally
symmetric solution of the vacuum Einstein equations with a negative
cosmological constant (for geometrical properties of AdS space and its uses
see, e.g., \cite{Grif09}). Quantum field theory in AdS background has long
been an active field of research for a variety of reasons. First of all, AdS
spacetime has a high degree of symmetry and, because of this, numerous
physical problems are exactly solvable in this geometry. The maximal
symmetry of AdS simplifies many aspects of the study of quantum fields and
the investigations of the corresponding field-theoretical effects may help
to develop the research tools and insights to deal with more complicated
geometries. Much of early interest to quantum field theory on AdS bulk was
motivated by principal questions of the quantization of fields on curved
backgrounds. The lack of global hyperbolicity and the presence of both
regular and irregular modes give rise to a number of new features which have
no analogues in quantum field theory on the Minkowski bulk. Being a constant
negative curvature manifold, AdS space provides a convenient infrared
regulator in interacting quantum field theories \cite{Call90}. Its natural
length scale can be used to regularize infrared divergences without reducing
the symmetries. The importance of this theoretical research was increased by
the natural appearance of AdS spacetime as a ground state in supergravity
and Kaluza-Klein theories and also as the near horizon geometry of the
extremal black holes and domain walls.

A further increase of interest is related to the crucial role of the AdS
geometry in two exciting developments of the past decade such as the AdS/CFT
correspondence and the braneworld scenario with large extra dimensions. The
AdS/CFT correspondence \cite{Mald98} (see also \cite{Ahar00}) represents a
realization of the holographic principle and relates string theories or
supergravity in the AdS bulk with a conformal field theory living on its
boundary. It has many interesting consequences and provides a powerful tool
for the investigation of gauge field theories. Among the recent developments
of the AdS/CFT correspondence is the application to strong-coupling problems
in condensed matter physics (familiar examples include holographic
superconductors, quantum phase transitions and topological insulators).
Moreover, the correspondence between the theories on AdS and Minkowski bulks
may be used to derive new results in mathematical physics, in particular, in
the theory of special functions (see, for instance, \cite{Bros12} and
references therein). The braneworld scenario (for reviews see \cite{Ruba01})
offers a new perspective on the hierarchy problem between the gravitational
and electroweak mass scales. The main idea to resolve the large hierarchy is
that the small coupling of four-dimensional gravity is generated by the
large physical volume of extra dimensions. Braneworlds naturally appear in
string/M-theory context and present intriguing possibilities to solve or to
address from a different point of view various problems in particle physics
and cosmology.

An inherent feature of all these models is that the boundary conditions on
the fields should be specified in order to completely determine the
dynamics. First of all the AdS spacetime is not globally hyperbolic and has
a time-like future null infinity. As a consequence of this, the information
may be lost to, or gained from, spatial infinity in finite coordinate time.
In order to define a consistent quantum field theory, appropriate boundary
conditions must be imposed \cite{Avis78,Brei82}. The general class of
allowed boundary conditions on the AdS boundary has been discussed in \cite%
{Ishi04}, based on the analysis of \cite{Ishi03}. Different boundary
conditions lead to different theories, in particular to different conformal
field theory duals. In braneworld models on AdS bulk, the presence of branes
gives rise to additional boundary conditions on the operator of a quantum
field. These conditions depend on the specific geometry of the brane and
have been discussed in \cite{Gher00,Flac01,Saha05} for models with $Z_{2}$
symmetry. Another type of boundary conditions is induced by the presence of
compact spatial dimensions. The extra compact dimensions are an inherent
feature of braneworld models arising from string and M-theories. In these
models one needs also to specify the periodicity conditions along compact
dimensions. Different conditions correspond to topologically inequivalent
field configurations \cite{Isha78}. The nontrivial periodicity conditions
lead to a number of interesting quantum field-theoretical effects, which
include instabilities in interacting field theories, topological mass
generation, and symmetry breaking.

All these types of boundary conditions, imposed on the field operator,
modify the spectrum of the zero-point fluctuations of a quantum field. As a
consequence, the vacuum expectation values (VEVs) of physical observables
are shifted by an amount depending on the geometry of the boundary and on
the type of the boundary condition. This general phenomena is known as the
Casimir effect. It has important implications on all scales, from mesoscopic
condensed matter physics to cosmology. Since the original research by
Casimir \cite{Casi48}, for the electromagnetic field in the geometry of two
parallel conducting plates, many theoretical and experimental works have
been done in this direction (for reviews see \cite{Most97}). In braneworlds,
the boundary conditions imposed on the bulk fields will give Casimir-type
contributions to the vacuum energy and to the vacuum forces acting on the
branes. The latter provide a natural mechanism for stabilizing the
interbrane distance (radion) in Randall-Sundrum-type models. The Casimir
energy gives a contribution to both the brane and the bulk cosmological
constants and should be taken into account in the self-consistent
formulation of the scenario. Motivated by these issues, the investigations
of the Casimir energy and related forces on AdS bulk have attracted a great
deal of attention (see, for instance, the references in \cite{Eliz13}). The
Casimir effect in higher-dimensional generalizations of the AdS spacetime
with compact internal spaces has been discussed in \cite{Flac03}.

An important physical characteristic of the vacuum state for charged fields
is the expectation value of the current density. It carries information
about the geometry and topology of the background space and is responsible
for the backreaction of the quantum field, as a source in semiclassical
Maxwell's equations. In the present paper we investigate the VEV of the
current density for a charged scalar field in background of locally AdS
spacetime with an arbitrary number of toroidally compactified spatial
dimensions, in the presence of a brane parallel to the AdS boundary. The
corresponding problem in the absence of the brane has been considered
recently in \cite{Beze15} and here we shall be mainly concerned with the
brane-induced effects. Both the zero and finite temperature expectation
values of the current density for charged scalar and fermionic fields in
background of flat spacetime with toroidal dimensions were investigated in \cite%
{Beze13c,Bell10}. Applications were given to the electronic subsystem of
cylindrical and toroidal carbon nanotubes described in terms of a $(2+1)$%
-dimensional effective field theory. The vacuum current densities for
charged scalar and Dirac spinor fields in de Sitter spacetime with
toroidally compact spatial dimensions are considered in \cite{Bell13b}. The
influence of boundaries on the vacuum currents in topologically nontrivial
spaces are studied in \cite{Bell13,Bell15} for scalar and fermionic fields.
The effects of nontrivial topology induced by the compactification of a
cosmic string along its axis have been discussed in \cite{Beze13}.

The organization of the paper is as follows. In the next section we specify
the bulk and boundary geometries under consideration and evaluate the
Hadamard function for a charged massive scalar field in both regions on the
right and on the left to the brane (referred to as R- and L-regions,
respectively). The brane-induced contributions are manifestly extracted and
they are presented in the form well suited for the investigation of the VEVs
of local physical observables bilinear in the field. As such an observable,
in section \ref{sec:Curr}, we consider the current density. The
corresponding VEVs are decomposed into boundary-free and brane-induced
contributions for both R- and L-regions. The asymptotic behavior of the
brane-induced contributions is considered near the brane, near the AdS
boundary and near the horizon. Limiting expressions are derived for small
and large proper lengths of compact dimensions. The main results are
summarized in section \ref{sec:Conc}. In Appendix we provide alternative
representations for the Hadamard functions in the R- and L-regions. The
expressions for the current densities obtained from these representations
are used for the investigation of the near brane asymptotic.

\section{Geometry of the problem and two-point functions}

\label{sec:Hadam}

\subsection{Set-up}

We consider a charged quantum scalar field $\varphi (x)$ with the mass $m$
and with the curvature coupling parameter $\xi $. In the presence of an
external classical gauge field $A_{\mu }$, the corresponding field equation
reads
\begin{equation}
\left( g^{\mu \nu }D_{\mu }D_{\nu }+m^{2}+\xi R\right) \varphi (x)=0,
\label{fieldeq}
\end{equation}%
where $D_{\mu }=\nabla _{\mu }+ieA_{\mu }$, $e$ is the charge of the field
quanta, $\nabla _{\mu }$ is the operator of the covariant derivative
associated with the metric tensor $g_{\mu \nu }$, and $R$ is the Ricci
scalar for the background spacetime. The background geometry in the present
paper is given by the interval
\begin{equation}
ds^{2}=g_{\mu \nu }dx^{\mu }dx^{\nu }=e^{-2y/a}\eta _{ik}dx^{i}dx^{k}-dy^{2},
\label{metric}
\end{equation}%
where $\eta _{ik}=\mathrm{diag}(1,-1,\ldots ,-1)$ is the metric tensor for $%
D $-dimensional Minkowski spacetime, $i,k=0,\ldots ,D-1$, and $\mu ,\nu $
run from 0 to $D$. The local geometry described by (\ref{metric}) coincides
with that for $(D+1)$-dimensional AdS spacetime of the radius $a$, expressed
in Poincar\'{e} coordinates. The corresponding metric tensor is a solution
of the Einstein equations with a negative cosmological constant $\Lambda
=-D(D-1)a^{-2}/2$ and for the Ricci scalar one has $R=-D(D+1)/a^{2}$.

The spatial topology considered here will be different from that for AdS.
Namely, we assume that the subspace with the coordinates $x^{l}$, $%
l=p+1,\ldots ,D-1$, is compactified to a $q$-dimensional torus $T^{q}$ with $%
q=D-p-1$ (for a recent review of quantum field-theoretical effects in
toroidal topology see \cite{Khan14}). We shall denote by $L_{l}$ the length
of the $l$th compact dimension, $0\leqslant x^{l}\leqslant L_{l}$. The
ranges of the remaining coordinates are $-\infty <x^{l}<+\infty $, $%
l=1,2,\ldots ,p$, and $-\infty <y<+\infty $. Consequently, for the subspace
perpendicular to the $y$-axis we take the topology $R^{p}\times T^{q}$. Note
that $L_{l}$ is the coordinate length of the compact dimension. For a fixed $%
y$, the proper length of the $l$th compact dimension is given by $%
L_{(p)l}=e^{-y/a}L_{l}$ and it decreases with increasing $y$. The coordinate
transformation
\begin{equation}
z=ae^{y/a},\;0\leqslant z<\infty ,  \label{z}
\end{equation}%
brings the interval (\ref{metric}) into manifestly conformally flat form
with the conformal factor $(a/z)^{2}$:%
\begin{equation}
ds^{2}=(a/z)^{2}(\eta _{ik}dx^{i}dx^{k}-dz^{2}).  \label{metric2}
\end{equation}%
In terms of the new coordinate $z$, the AdS boundary and horizon are
presented by the hypersurfaces $z=0$ and $z=\infty $, respectively. For the
proper length, measured by an observer with a fixed coordinate $z$, one gets
$L_{(p)l}=aL_{l}/z$.

We consider a field theory in a non-simply connected spacetime and for the
complete formulation of the problem the periodicity conditions along compact
dimensions should be specified. Here we impose the conditions%
\begin{equation}
\varphi (t,x^{1},\ldots ,x^{l}+L_{l},\ldots ,y)=e^{i\alpha _{l}}\varphi
(t,x^{1},\ldots ,x^{l},\ldots ,y),  \label{PerC}
\end{equation}%
for $l=p+1,\ldots ,D-1$ with constant phases $\alpha _{l}$. In the
literature, the most frequently considered special cases correspond to $%
\alpha _{l}=0$ (untwisted scalar) and $\alpha _{l}=\pi $ (twisted scalar).
As we will see below, the nontrivial phases in the periodicity conditions
give rise to the vacuum currents along compact dimensions. In the discussion
below we shall assume that the gauge field $A_{\mu }$ is constant. Though
the corresponding field strength vanishes, because of the nontrivial
topology of the background spacetime, the VEVs of physical observables will
be influenced by this sort of field configuration. This is an Aharonov-Bohm
like effect of a constant gauge field. Let us consider two sets of the
fields $(\varphi (x),A_{\mu })$ and $(\varphi ^{\prime }(x),A_{\mu }^{\prime
})$ connected by the gauge transformation $\varphi (x)=e^{-ie\chi
(x)}\varphi ^{\prime }(x)$, $A_{\mu }=A_{\mu }^{\prime }+\partial _{\mu
}\chi (x)$. For a constant gauge field, taking the function $\chi (x)=A_{\mu
}x^{\mu }$, we see that in the new gauge the vector potential vanishes, $%
A_{\mu }^{\prime }=0$. But, after the gauge transformation the vector
potential of the former gauge appears in the periodicity conditions for the
new field operator:%
\begin{equation}
\varphi ^{\prime }(t,x^{1},\ldots ,x^{l}+L_{l},\ldots ,y)=e^{i\tilde{\alpha}%
_{l}}\varphi ^{\prime }(t,x^{1},\ldots ,x^{l},\ldots ,y),  \label{PerC2}
\end{equation}%
with $l=p+1,\ldots ,D-1$, and with the new phases%
\begin{equation}
\tilde{\alpha}_{l}=\alpha _{l}+eA_{l}L_{l}.  \label{alfl}
\end{equation}%
In particular, nontrivial phases are generated for untwisted and twisted
scalars. The phase shift in (\ref{PerC2}) is related to the magnetic flux $%
\Phi _{l}$ enclosed by the $l$th compact dimension: $eA_{l}L_{l}=-2\pi \Phi
_{l}/\Phi _{0}$, where $\Phi _{0}=2\pi /e$ is the flux quantum. Note that
the gauge field fluxes play an important role in recent developments of
string theory compactifications (for a review see \cite{Doug07}).

\begin{figure}[tbph]
\begin{center}
\epsfig{figure=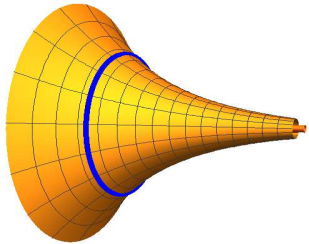,width=5cm,height=4cm}
\end{center}
\caption{The spatial section of the geometry for $D=2$ embedded into a
3-dimensional Euclidean space. The thick circle corresponds to the brane.}
\label{fig1}
\end{figure}

In the problem under consideration, the second type of boundary condition
imposed on the field operator is induced by a brane parallel to the AdS
boundary and located at $y=y_{0}$. The corresponding value for the conformal
coordinate $z$ we shall denote by $z_{0}=ae^{y_{0}/a}$. On the brane we
assume a gauge invariant boundary condition of the Robin type:
\begin{equation}
(1+\beta n^{\mu }D_{\mu })\varphi (x)=0,\quad y=y_{0},  \label{Rob}
\end{equation}%
where $\beta $ is a constant and $n^{\mu }$ is the inward pointing normal to
the brane. For the latter one has $n^{\mu }=\delta _{D}^{\mu }$ in the
region $y>y_{0}$ and $n^{\mu }=-\delta _{D}^{\mu }$ in the region $y<y_{0}$.
Note that, in general, the value of the coefficient $\beta $ for these
regions could be different. The Robin boundary condition is a generalization
of Dirichlet and Neumann conditions and naturally appears in a number of
physical problems, including those in braneworld scenario (see below). The
spatial geometry under consideration for $D=2$, embedded into the
3-dimensional Euclidean space, is displayed in figure \ref{fig1}. The
compact dimension is presented by the circles and the thick circle
corresponds to the location of the brane. We have also depicted the gauge
field flux tube enclosed by the compact dimension. The proper length of the
compact dimension decreases with increasing $y$.

The physical quantity we are interested in is the VEV of the current
density, $\langle 0|j_{\mu }(x)|0\rangle \equiv \langle j_{\mu }(x)\rangle $%
, for the field $\varphi (x)$, where $|0\rangle $ stands for the vacuum
state. In quantum field theory on curved backgrounds the choice of the
vacuum state is among the basic points. In what follows it will be assumed
that the field is prepared in the Poincar\'{e} vacuum state. The latter is
realized by the mode functions of the field which are obtained by solving
the field equation in Poincar\'{e} coordinates (for the discussion of the
relation between the Poincar\'{e} and global vacua see, for instance, \cite%
{Dani99}). The operator of the current density for a charged scalar field is
defined by the expression
\begin{equation}
j_{\mu }(x)=ie[\varphi ^{+}(x)D_{\mu }\varphi (x)-(D_{\mu }\varphi
^{+}(x))\varphi (x)].  \label{jmu}
\end{equation}%
Its VEV is among the most important characteristics of the vacuum state. The
procedure we shall use here for the evaluation of the expectation value is
based on the formula
\begin{equation}
\langle j_{\mu }(x)\rangle =\frac{i}{2}e\lim_{x^{\prime }\rightarrow
x}(\partial _{\mu }-\partial _{\mu }^{\prime }+2ieA_{\mu })G(x,x^{\prime }),
\label{jl1}
\end{equation}%
where
\begin{eqnarray}
G(x,x^{\prime }) &=&\langle 0|\varphi (x)\varphi ^{+}(x^{\prime })+\varphi
^{+}(x^{\prime })\varphi (x)|0\rangle  \notag \\
&=&\sum_{\sigma }\sum_{s=\pm }\varphi _{\sigma }^{(s)}(x)\varphi _{\sigma
}^{(s)\ast }(x^{\prime }).  \label{Had}
\end{eqnarray}%
is the Hadamard function for the vacuum state under consideration. In (\ref%
{Had}), the summation goes over a complete orthonormal set of positive- and
negative-energy mode functions $\varphi _{\sigma }^{(\pm )}(x)$ specified by
a collective index $\sigma $ involving the corresponding quantum numbers.
The mode functions obey the quasiperiodicity conditions (\ref{PerC}) and the
condition (\ref{Rob}) on the brane.

So, as the first step, we shall evaluate the Hadamard function. Though the
background AdS spacetime is homogeneous, the brane at $y=y_{0}$ has nonzero
extrinsic curvature tensor and its sides are not equivalent. In particular,
the VEVs differ in the regions on the right and on the left of the brane.
The consideration requires different procedures for these regions and we
discuss them separately. In what follows the regions $y>y_{0}$ and $y<y_{0}$
will be referred to as R-region (right region) and L-region (left region),
respectively.

\subsection{Hadamard function in the R-region}

We shall work in the gauge with the fields $(\varphi ^{\prime }(x),A_{\mu
}^{\prime }=0)$, omitting the prime. In accordance with the problem
symmetry, the mode functions can be factorized as
\begin{equation}
\varphi _{\sigma }^{(\pm )}(x)=z^{D/2}Z_{\nu }(\lambda z)e^{ik_{r}x^{r}\mp
i\omega t},  \label{Modes}
\end{equation}%
where the summation over $r$ in the exponent goes for $r=1,\ldots ,D-1$, $%
Z_{\nu }(x)$ is a cylinder function of the order%
\begin{equation}
\nu =\sqrt{D^{2}/4-D(D+1)\xi +m^{2}a^{2}},  \label{nu}
\end{equation}%
and%
\begin{equation}
\omega =\sqrt{\lambda ^{2}+k^{2}},\;k^{2}=\sum_{l=1}^{D-1}k_{l}^{2}.
\label{lamb}
\end{equation}%
For imaginary values of the order $\nu $ the vacuum state becomes unstable
\cite{Brei82} and in what follows we shall assume the values of the
parameters for which $\nu \geqslant 0$. In the cases of conformally and
minimally coupled massless fields one has $\nu =1/2$ and $\nu =D/2$,
respectively. In the former case, we have the standard conformal relation
with the modes in the problem on Minkowski bulk with toroidal dimensions. In (%
\ref{Modes}), for the components of the momentum one has $-\infty
<k_{l}<+\infty $, $l=1,\ldots ,p$, and
\begin{equation}
k_{l}=(2\pi n_{l}+\tilde{\alpha}_{l})/L_{l},\;l=p+1,\ldots ,D-1,  \label{kl}
\end{equation}%
with $n_{l}=0,\pm 1,\pm 2,\ldots $. The eigenvalues (\ref{kl}) for the
components along compact dimensions are directly obtained from the
quasiperiodicity conditions (\ref{PerC2}).

In the R-region, the function $Z_{\nu }(\lambda z)$ is a linear combination
of the Bessel and Neumann functions, $J_{\nu }(\lambda z)$ and $Y_{\nu
}(\lambda z)$. First we consider the case when for all the modes $\lambda $
is real. The changes in the evaluation procedure in the case when the modes
with purely imaginary values of $\lambda $ are allowed (bound states) will
be discussed below. The relative coefficient in the combination of the
Bessel and Neumann functions is determined from the boundary condition (\ref%
{Rob}) on the brane (with $D_{\mu }=\partial _{\mu }$) and one gets%
\begin{equation}
Z_{\nu }(\lambda z)=C_{\sigma }g_{\nu }(\lambda z_{0},\lambda z),  \label{Z}
\end{equation}%
where, for the further convenience, we have introduced the function%
\begin{equation}
g_{\nu }(u,v)=J_{\nu }(v)\bar{Y}_{\nu }(u)-\bar{J}_{\nu }(u)Y_{\nu }(v).
\label{gnu}
\end{equation}%
Here, for a given function $F(x)$, we use the notation
\begin{equation}
\bar{F}(x)=A_{0}F(x)+B_{0}xF^{\prime }(x),  \label{Fbar}
\end{equation}%
with the coefficients
\begin{equation}
A_{0}=1+\frac{D}{2}\delta _{y}\beta /a,\quad B_{0}=\delta _{y}\beta /a.
\label{A0}
\end{equation}%
In (\ref{A0}) and in what follows, $\delta _{y}=1$ in the R-region and $%
\delta _{y}=-1$ in the L-region.

Now, the set of quantum numbers $\sigma $ is specified by $\sigma =(\lambda ,%
\mathbf{k}_{p},\mathbf{n}_{q})$, where $\mathbf{k}_{p}=(k_{1},\ldots ,k_{p})$
is the momentum in non-compact space and $\mathbf{n}_{q}=(n_{p+1},\ldots
,n_{D-1})$. The coefficient $C_{\sigma }$ is determined from the
normalization condition
\begin{equation}
\int d^{D}x\,g^{00}\sqrt{|g|}\varphi _{\sigma }^{(s)}(x)\varphi _{\sigma
^{\prime }}^{(s^{\prime })\ast }(x)=\frac{\delta _{ss^{\prime }}}{2\omega }%
\delta (\lambda -\lambda ^{\prime })\delta (\mathbf{k}_{p}-\mathbf{k}%
_{p}^{\prime })\delta _{\mathbf{n}_{q},\mathbf{n}_{q}^{\prime }},
\label{Norm}
\end{equation}%
where the $y$-integration goes over $[y_{0},\infty )$. By using the mode
functions (\ref{Modes}) with the radial function from (\ref{Z}), we find%
\begin{equation}
|C_{\sigma }|^{2}=\frac{a^{1-D}\lambda }{2\left( 2\pi \right) ^{p}\omega
V_{q}}\left[ \bar{J}_{\nu }^{2}(\lambda z_{0})+\bar{Y}_{\nu }^{2}(\lambda
z_{0})\right] ^{-1},  \label{Csig}
\end{equation}%
with $V_{q}=L_{p+1}\cdots L_{D-1}$ being the volume of the compact subspace.

Having determined the complete set of normalized modes, from the mode-sum in
(\ref{Had}), for the Hadamard function one obtains the representation%
\begin{eqnarray}
G(x,x^{\prime }) &=&\frac{\left( zz^{\prime }\right) ^{D/2}}{\left( 2\pi
\right) ^{p}a^{D-1}V_{q}}\sum_{\mathbf{n}_{q}}\int d\mathbf{k}%
_{p}\,e^{ik_{r}\Delta x^{r}}\int_{0}^{\infty }d\lambda \,\lambda  \notag \\
&&\times \frac{\cos (\Delta t\sqrt{\lambda ^{2}+k^{2}})}{\sqrt{\lambda
^{2}+k^{2}}}\frac{g_{\nu }(\lambda z_{0},\lambda z)g_{\nu }(\lambda
z_{0},\lambda z^{\prime })}{\bar{J}_{\nu }^{2}(\lambda z_{0})+\bar{Y}_{\nu
}^{2}(\lambda z_{0})}.  \label{GL1}
\end{eqnarray}%
In order to extract explicitly the brane-induced contribution, we subtract
from (\ref{GL1}) the corresponding Hadamard function in the boundary-free
AdS background which will be denoted by $G_{0}(x,x^{\prime })$. The latter
is obtained from (\ref{GL1}) replacing the last fraction in the right-hand
side by the product $J_{\nu }(\lambda z)J_{\nu }(\lambda z^{\prime })$. By
using the identity%
\begin{equation}
\frac{g_{\nu }(\lambda z_{0},\lambda z)g_{\nu }(\lambda z_{0},\lambda
z^{\prime })}{\bar{J}_{\nu }^{2}(\lambda z_{0})+\bar{Y}_{\nu }^{2}(\lambda
z_{0})}=J_{\nu }(\lambda z)J_{\nu }(\lambda z^{\prime })-\frac{1}{2}%
\sum_{j=1,2}\frac{\bar{J}_{\nu }(\lambda z_{0})}{\bar{H}_{\nu
}^{(j)}(\lambda z_{0})}H_{\nu }^{(j)}(\lambda z)H_{\nu }^{(j)}(\lambda
z^{\prime }),  \label{Ident}
\end{equation}%
with $H_{\nu }^{(1,2)}(x)$ being the Hankel functions, the following
decomposition is obtained%
\begin{eqnarray}
G(x,x^{\prime }) &=&G_{0}(x,x^{\prime })-\frac{a^{1-D}\left( zz^{\prime
}\right) ^{D/2}}{2\left( 2\pi \right) ^{p}V_{q}}\sum_{\mathbf{n}_{q}}\int d%
\mathbf{k}_{p}\,e^{ik_{r}\Delta x^{r}}\int_{0}^{\infty }d\lambda  \notag \\
&&\times \lambda \frac{\cos (\Delta t\sqrt{\lambda ^{2}+k^{2}})}{\sqrt{%
\lambda ^{2}+k^{2}}}\sum_{j=1,2}\frac{\bar{J}_{\nu }(\lambda z_{0})}{\bar{H}%
_{\nu }^{(j)}(\lambda z_{0})}H_{\nu }^{(j)}(\lambda z)H_{\nu }^{(j)}(\lambda
z^{\prime }).  \label{GL2}
\end{eqnarray}%
Now, assuming that $z>z_{0}$, we rotate the integration contour of the last
integral by the angle $\pi /2$ for $j=1$ and by $-\pi /2$ for $j=2$.
Introducing the modified Bessel functions $I_{\nu }(x)$ and $K_{\nu }(x)$,
the following final expression is obtained:%
\begin{eqnarray}
G(x,x^{\prime }) &=&G_{0}(x,x^{\prime })-\frac{4\left( zz^{\prime }\right)
^{D/2}}{\left( 2\pi \right) ^{p+1}a^{D-1}V_{q}}\sum_{\mathbf{n}_{q}}\int d%
\mathbf{k}_{p}\,e^{ik_{r}\Delta x^{r}}\,  \notag \\
&&\times \int_{k}^{\infty }du\,u\frac{\cosh (\Delta t\sqrt{u^{2}-k^{2}})}{%
\sqrt{u^{2}-k^{2}}}\frac{\bar{I}_{\nu }(uz_{0})}{\bar{K}_{\nu }(uz_{0})}%
K_{\nu }(uz)K_{\nu }(uz^{\prime }),  \label{GL3}
\end{eqnarray}%
where the notation with overbar is defined by (\ref{Fbar}), (\ref{A0}) with $%
\delta _{y}=1$.

In deriving (\ref{GL3}) we have assumed that for all the modes of the field $%
\lambda $ is real. In addition to these modes, bound states can be present.
For them $\lambda $ is purely imaginary, $\lambda =i\eta $, $\eta >0$, and
the mode functions have the form
\begin{equation}
\varphi _{\sigma }^{(\pm )}(x)=C_{\sigma }^{(\mathrm{b})}z^{D/2}K_{\nu
}(\eta z)e^{ik_{r}x^{r}\mp i\omega (\eta )t},  \label{BS}
\end{equation}%
where $\omega (\eta )=\sqrt{k^{2}-\eta ^{2}}$. Let us denote by $%
k_{(q)}^{(0)}$ the lowest value of the momentum in the compact subspace.
Assuming that $|\tilde{\alpha}_{i}|\leqslant \pi $, one has
\begin{equation}
k_{(q)}^{(0)2}=\sum_{i=p+1}^{D-1}\tilde{\alpha}_{i}^{2}/L_{i}^{2}.
\label{kq0}
\end{equation}%
If $\eta >k_{(q)}^{(0)}$, then there are modes for which the energy is
purely imaginary and the vacuum state is unstable. In order to have a stable
vacuum, in what follows we assume that $\eta <k_{(q)}^{(0)}$. From the
boundary condition (\ref{Rob}) it follows that for bound states the possible
values of $\eta $ are roots of the equation%
\begin{equation}
\bar{K}_{\nu }(\eta z_{0})=0.  \label{BSval}
\end{equation}%
By using the recurrence relation for the Macdonald function, this equation
can be rewritten as%
\begin{equation}
\left( \nu -D/2-a/\beta \right) K_{\nu }(u)+uK_{\nu -1}(u)=0,  \label{BSval2}
\end{equation}%
with $u=\eta z_{0}$. From here it follows that there are no bound states for
$a/\beta <\nu -D/2$ (for the special mode in the case $a/\beta =\nu -D/2$
see below). For $a/\beta >\nu -D/2$ a single bound state $\lambda =i\eta $
appears. The corresponding root $\eta $ increases with increasing $a/\beta $
and for some critical value $\beta =\beta _{\mathrm{R}}=\beta _{\mathrm{R}%
}(k_{(q)}^{(0)})$ one gets $\eta =k_{(q)}^{(0)}$. Here, $\beta _{\mathrm{R}}$
is the value of the Robin coefficient for which the root of the equation is
equal to $k_{(q)}^{(0)}$. The stability of the vacuum state requires the
condition $1/\beta <1/\beta _{\mathrm{R}}$. The critical value for the Robin
coefficient depends on the lengths of the compact dimensions, on the phases
in periodicity conditions and on the mass of the field through the parameter
$\nu $. Note that for a brane on AdS bulk with all dimensions being
non-compact one has $k_{(q)}^{(0)}=0$ and all the modes with $\lambda =i\eta
$, $\eta >0$, lead to the instability. Hence, in models with compact
dimensions the stability condition, in general, is less restrictive.
Assuming that $a/\beta >\nu -D/2$, let us denote by $u=u_{\nu
}^{(R)}(a/\beta )$ the root of the equation (\ref{BSval2}). This root
increases with increasing $a/\beta $ and does not depend on the location of
the brane. The stability condition for the vacuum state is written as $%
u_{\nu }^{(R)}(a/\beta )<k_{(q)}^{(0)}z_{0}$. From here it follows that, for
fixed values of the other parameters, when the brane approaches the AdS
boundary ($z_{0}$ decreases), started from the critical value $z_{0}=u_{\nu
}^{(R)}(a/\beta )/k_{(q)}^{(0)}$, the vacuum in the R-region becomes
unstable.

The coefficient $C_{\sigma }^{(\mathrm{b})}$ in (\ref{BS}) is found from the
normalization condition (\ref{Norm}) making the replacement $\delta (\lambda
-\lambda ^{\prime })\rightarrow \delta _{\eta \eta ^{\prime }}$. By using
the result for the integral involving the square of the Macdonald function
\cite{Prud86}, one gets%
\begin{equation}
|C_{\sigma }^{(\mathrm{b})}|^{2}=-\frac{\left( 2\pi \right) ^{-p}a^{1-D}\eta
\bar{I}_{\nu }(\eta z_{0})}{V_{q}\omega (\eta )z_{0}\bar{K}_{\nu }^{\prime
}(\eta z_{0})}.  \label{Cbsig}
\end{equation}%
In deriving this expression we have used the relations
\begin{equation}
K_{\nu }(x)=\frac{B_{0}}{\bar{I}_{\nu }(x)},\;B_{0}^{2}(x^{2}+\nu
^{2})-A_{0}^{2}=B_{0}x\frac{\bar{K}_{\nu }^{\prime }(x)}{K_{\nu }(x)},
\label{RelK}
\end{equation}%
valid for $x=\eta z_{0}$, with $\eta z_{0}$ being the solution of (\ref%
{BSval}). These relations are obtained by making use of (\ref{BSval}) and
the Wronskian relation for the modified Bessel functions.

Now, in the presence of the bound state, the mode sum for the Hadamard
function has two contributions. The first one comes from the modes with real
$\lambda $ and is given by the expression (\ref{GL1}). The second
contribution comes from the bound state. For the latter, by using the
corresponding mode functions (\ref{BS}) and the normalization coefficient
from (\ref{Cbsig}), one finds%
\begin{eqnarray}
G^{(\mathrm{b})}(x,x^{\prime }) &=&-\frac{2\alpha ^{1-D}(zz^{\prime })^{%
\frac{D}{2}}}{\left( 2\pi \right) ^{p}V_{q}z_{0}}\sum_{\mathbf{n}_{q}}\int d%
\mathbf{k}_{p}\,e^{ik_{r}\Delta x^{r}}\eta  \notag \\
&&\times \frac{\bar{I}_{\nu }(\eta z_{0})}{\bar{K}_{\nu }^{\prime }(\eta
z_{0})}\frac{\cos (\omega (\eta )\Delta t)}{\omega (\eta )}K_{\nu }(\eta
z)K_{\nu }(\eta z^{\prime }).  \label{Gb}
\end{eqnarray}%
The evaluation of the part coming from the modes with real $\lambda $ is
similar to that we have described above. The difference arises in the step
when we rotate the integration contour in (\ref{GL2}). Now, the integrand in
this expression has poles $\lambda =\pm i\eta $ on the imaginary axis, where
$\eta $ is the root of (\ref{BSval}). After the rotation, the integration
contour has to pass round these poles on the right by small semicircles. The
integrals over the semicircles around $\lambda =i\eta $ and $\lambda =-i\eta
$ give the residue at $\lambda =i\eta $ multiplied by $2\pi i$. It can be
seen that this residue term exactly cancels the contribution of the bound
state in (\ref{Gb}). Hence, we conclude that the expression (\ref{GL3}) for
the Hadamard function is valid in the case of the presence of bound states
as well.

In addition to the modes discussed above, a mode may be present for which $%
\lambda =0$ and $\omega =k$. \ For this mode the function $Z_{\nu }$ in (\ref%
{Modes}) is a linear combination of $z^{\nu }$ and $z^{-\nu }$. The part
with $z^{\nu }$ is excluded by the normalizability condition and the mode
functions have the form%
\begin{equation}
\varphi _{(\mathrm{R})\sigma }^{(\pm )}(x)=C_{(\mathrm{R})}z^{D/2-\nu
}e^{ik_{r}x^{r}\mp ikt}.  \label{ModeSp}
\end{equation}%
These modes are normalizable under the condition $\nu >1$ and for the
coefficient one finds%
\begin{equation}
C_{(\mathrm{R})}^{2}=\frac{\left( \nu -1\right) z_{0}^{2\nu -2}}{\left( 2\pi
\right) ^{p}V_{q}a^{D-1}k}.  \label{Csp}
\end{equation}%
From the boundary condition on the brane it follows that the mode is allowed
for the special value of the Robin coefficient determined from
\begin{equation}
\beta /a=1/(\nu -D/2).  \label{betsp}
\end{equation}%
For $\nu =D/2$ this value corresponds to Neumann boundary condition and in
this case the mode function does not depend on the coordinate $z$. An
example of this special case is realized by a minimally coupled massless
scalar field. Note that this special mode for a scalar field is the analog
of the zero mode of the graviton in Randall-Sundrum 1-brane model \cite%
{Rand99}.

For $\nu >1$ and in the case of Robin boundary condition with (\ref{betsp}),
the contribution of the special mode (\ref{ModeSp}) should be separately
added to the Hadamard function in formulas (\ref{GL1}) and (\ref{GL2}) (but
not to (\ref{GL3}), see below). Note that the mode function (\ref{ModeSp})
can be written as $\varphi _{(\mathrm{R})\sigma }^{(\pm )}(x)=\Omega _{(%
\mathrm{R})}(z)\varphi _{(\mathrm{M})\sigma }^{(\pm )}(x)$, where%
\begin{equation}
\Omega _{(\mathrm{R})}(z)=\sqrt{2\left( \nu -1\right) }z_{0}^{\nu -1}\frac{%
z^{D/2-\nu }}{a^{(D-1)/2}},  \label{Omz}
\end{equation}%
and $\varphi _{(\mathrm{M})\sigma }^{(\pm )}(x)$ are the mode functions for
a massless scalar field in $D$-dimensional Minkowski spacetime with the
spatial topology $R^{p}\times T^{q}$. From here it follows that the
contribution of the mode (\ref{ModeSp}) to the Hadamard function is
expressed as $G_{(\mathrm{R})}(x,x^{\prime })=\Omega _{(\mathrm{R}%
)}(z)\Omega _{(\mathrm{R})}(z^{\prime })G_{R^{p}\times T^{q}}^{(\mathrm{M}%
)}(x,x^{\prime })$, where $G_{R^{p}\times T^{q}}^{(\mathrm{M})}(x,x^{\prime
})$ is the corresponding function for a massless scalar field in $D$%
-dimensional Minkowski spacetime with the spatial topology $R^{p}\times
T^{q} $. By using the expression for the latter one gets%
\begin{equation}
G_{(\mathrm{R})}(x,x^{\prime })=\frac{2\Omega _{(\mathrm{R})}(z)\Omega _{(%
\mathrm{R})}(z^{\prime })}{\left( 2\pi \right) ^{p+1/2}V_{q}}\sum_{\mathbf{n}%
_{q}}e^{ik_{l}\Delta x^{l}}\frac{K_{(p-1)/2}(k_{(q)}s_{p})}{%
(s_{p}/k_{(q)})^{(p-1)/2}}.  \label{GRs}
\end{equation}%
where the summation in the exponent goes over $l=p+1,\ldots ,D-1$. In this
expression we have defined $s_{p}=\sqrt{|\Delta \mathbf{x}_{p}|^{2}-\left(
\Delta t\right) ^{2}}$ and
\begin{equation}
k_{(q)}^{2}=\sum_{l=p+1}^{D-1}k_{l}^{2}=\sum_{l=p+1}^{D-1}(2\pi n_{l}+\tilde{%
\alpha}_{l})^{2}/L_{l}^{2},  \label{kq}
\end{equation}%
is the squared momentum in the compact subspace.

It can be seen that the expression (\ref{GL3}) for the Hadamard function is
not changed by the presence of the mode (\ref{ModeSp}). Indeed, under the
condition (\ref{betsp}), the contribution (\ref{GRs}) of this mode is
separately added to the right-hand sides of (\ref{GL1}) and (\ref{GL2}). Now
we should take into account that the integrand in (\ref{GL2}) with separate $%
j$ has a simple pole at $\lambda =0$ and in the rotation of the integration
contour this pole should be avoided by arcs of a circle of small radius. The
contributions of the integrals over these arcs exactly cancel the
contribution (\ref{GRs}) of the special mode and, as a consequence of this,
the representation (\ref{GL3}) is not changed.

\subsection{Hadamard function in the L-region}

In the region between the brane and AdS boundary, $y<y_{0}$, the mode
functions still have the form (\ref{Modes}). From the normalizability
condition it follows that for $\nu \geqslant 1$ we should take
\begin{equation}
Z_{\nu }(\lambda z)=C_{\sigma }J_{\nu }(\lambda z).  \label{Zl}
\end{equation}%
For the part of the solution with the Neumann function, the normalization
integral diverges on the AdS boundary $z=0$. In AdS/CFT correspondence
normalizable and non-normalizable modes are dual to states and sources,
respectively. The key feature of AdS spacetime is the presence of a timelike
boundary at infinity where appropriate boundary conditions should be imposed
in order to have well-defined dynamics \cite{Avis78,Brei82}. In the problem
at hand, for the unique quantization procedure, in the range $0\leqslant \nu
<1$, we need to specify the boundary condition on the AdS boundary. For a
scalar field, the Dirichlet and Neumann boundary conditions are the most
frequently used ones. The general class of allowed boundary conditions of
the Robin type, has been discussed in \cite{Ishi04}. Here, we choose (\ref%
{Zl}) for all values of $\nu \geqslant 0$ that corresponds to Dirichlet
condition on the AdS boundary in the case $0\leqslant \nu <1$ (note that the
analytic continuation to the Euclidean section automatically selects this
boundary condition \cite{Camp91}).

From the boundary condition on the brane at $y=y_{0}$ it follows that the
eigenvalues of $\lambda $ are roots of the equation
\begin{equation}
\bar{J}_{\nu }(\lambda z_{0})=0,  \label{Jzer}
\end{equation}%
where the notation with overbar is defined in accordance with (\ref{Fbar})
where now in (\ref{A0}) $\delta _{y}=-1$. Hence, in the L-region the
spectrum for $\lambda $ is discrete. First we consider the case when all the
roots of (\ref{Jzer}) are real. Let us denote by $x=\gamma _{n}$, $%
n=1,2,\ldots $, the positive zeros of the functions $\bar{J}_{\nu }(x)$.
Then, for the eigenvalues of $\lambda $ one has $\lambda =\gamma _{n}/z_{0}$
(for a mode with purely imaginary $\lambda $ see below). The normalization
coefficient is determined from (\ref{Norm}), with the replacement $\delta
(\lambda -\lambda ^{\prime })\rightarrow \delta _{nn^{\prime }}$ and with
the $y$-integration over $(-\infty ,y_{0}]$. By using the standard result
for the integral involving the square of the Bessel function \cite{Prud86},
we find%
\begin{equation}
|C_{\sigma }|^{2}=\frac{\lambda z_{0}T_{\nu }(\lambda z_{0})}{\left( 2\pi
\right) ^{p}a^{D-1}V_{q}z_{0}^{2}\omega },  \label{CR}
\end{equation}%
with the notation%
\begin{equation}
T_{\nu }(x)=x[x^{2}J_{\nu }^{\prime 2}(x)+(x^{2}-\nu ^{2})J_{\nu
}^{2}(x)]^{-1}.  \label{T}
\end{equation}%
Note that in the latter expression we could substitute $xJ_{\nu }^{\prime
}(x)=-A_{0}J_{\nu }(x)/B_{0}$.

Plugging the mode functions into the mode-sum (\ref{Had}), for the Hadamard
function we get the expression%
\begin{eqnarray}
G(x,x^{\prime }) &=&\frac{2a^{1-D}\left( zz^{\prime }\right) ^{D/2}}{\left(
2\pi \right) ^{p}V_{q}z_{0}^{2}}\sum_{\mathbf{n}_{q}}\int d\mathbf{k}%
_{p}\,e^{ik_{r}\Delta x^{r}}\sum_{n}\,\gamma _{n}  \notag \\
&&\times \frac{\cos (\Delta t\sqrt{\gamma _{n}^{2}/z_{0}^{2}+k^{2}})}{\sqrt{%
\gamma _{n}^{2}/z_{0}^{2}+k^{2}}}T_{\nu }(\gamma _{n})J_{\nu }(\gamma
_{n}z/z_{0})J_{\nu }(\gamma _{n}z^{\prime }/z_{0}).  \label{GR}
\end{eqnarray}%
This expression involves the roots $\gamma _{n}$ which are given implicitly
and is not convenient for the further evaluation of the current density. In
order to obtain more workable expression we apply to the series over $n$ the
summation formula \cite{Saha87}
\begin{eqnarray}
\sum_{n=1}^{\infty }T_{\nu }(\gamma _{n})f(\gamma _{n}) &=&\frac{1}{2}%
\int_{0}^{\infty }du\,f(u)-\frac{1}{2\pi }\int_{0}^{\infty }du\,\frac{\bar{K}%
_{\nu }(u)}{\bar{I}_{\nu }(u)}  \notag \\
&&\times \left[ e^{-\nu \pi i}f(iu)+e^{\nu \pi i}f(-iu)\right] .
\label{sumAP}
\end{eqnarray}%
The part in the Hadamard function coming from the first term in the
right-hand side of (\ref{sumAP}) gives the corresponding function in the
geometry without the brane. As a result, we get the following decomposed
representation%
\begin{eqnarray}
G(x,x^{\prime }) &=&G_{0}(x,x^{\prime })-\frac{4a^{1-D}\left( zz^{\prime
}\right) ^{D/2}}{\left( 2\pi \right) ^{p+1}V_{q}}\sum_{\mathbf{n}_{q}}\int d%
\mathbf{k}_{p}\,e^{ik_{r}\Delta x^{r}}\int_{k}^{\infty }du  \notag \\
&&\times \lambda \frac{\cosh (\Delta t\sqrt{u^{2}-k^{2}})}{\sqrt{u^{2}-k^{2}}%
}\frac{\bar{K}_{\nu }(uz_{0})}{\bar{I}_{\nu }(uz_{0})}I_{\nu }(uz)I_{\nu
}(uz^{\prime }).  \label{GR2}
\end{eqnarray}%
Comparing with (\ref{GL3}), we see that the expressions for the brane-induced
parts in the Hadamard function in the R- and L-regions are obtained
from each other by the replacements $I_{\nu }(x)\rightleftarrows K_{\nu }(x)$
and with the replacement $\beta \rightarrow -\beta $ in the notations with
overbars. Comparing with (\ref{GR}), we see the important advantages of the
representation (\ref{GR2}): (i) the contribution of the brane is manifestly
separated, (ii) the explicit knowledge of the zeros $\gamma _{n}$ is not
required, (iii) the integrand in (\ref{GR2}) is monotonic instead of the
oscillatory behavior in (\ref{GR}) and (iv) the representation (\ref{GR2})
holds in the presence of the bound state and of the special mode (see
below) as well. In particular, the second and third points are important in the
numerical evaluation of the vacuum currents.

Depending on the value of the Robin coefficient $\beta $, the equation (\ref%
{Jzer}) can have purely imaginary roots $\lambda =i\eta $, $\eta >0$. In
this case, for the stability of the vacuum state we should assume that $\eta
<k_{(q)}^{(0)}$. From the boundary condition on the brane it follows that
the allowed values for $\eta $ are roots of the equation $\bar{I}_{\nu
}(\eta z_{0})=0$ which is written in the explicit form as
\begin{equation}
\left( D/2+\nu -a/\beta \right) I_{\nu }(u)+uI_{\nu +1}(u)=0,
\label{Immodes}
\end{equation}%
where $u=\eta z_{0}$. From here we conclude that the modes under
consideration are absent in the case $a/\beta <D/2+\nu $ (the special mode
for the case $a/\beta =D/2+\nu $ is discussed below). For $a/\beta >D/2+\nu $%
, the equation (\ref{Immodes}) has a single positive solution, $u=u_{\nu }^{(%
\mathrm{L})}(a/\beta )$, which increases with increasing $a/\beta $. Started
from the critical value of $a/\beta $, denoted here by $a/\beta _{\mathrm{L}%
} $ and determined from the condition $u_{\nu }^{(L)}(a/\beta _{\mathrm{L}%
})=k_{(q)}^{(0)}z_{0}$, the vacuum becomes unstable. Note that the critical
values $\beta _{\mathrm{R}}$ and $\beta _{\mathrm{L}}$ of the coefficient in
Robin boundary condition are different for the R- and L-region. As a result
of this, there are values of $\beta $ for which the vacuum is stable in the
one region and unstable in the other. We see that, under the condition $%
a/\beta >D/2+\nu $, when the location of the brane approaches the AdS
boundary, started from the critical value $z_{0}=u_{\nu }^{(\mathrm{L}%
)}(a/\beta )/k_{(q)}^{(0)}$, the vacuum state becomes unstable. The
summation formula (\ref{sumAP}) is valid also in the presence of purely
imaginary roots if we add to the left-hand side the contribution from the
corresponding modes. This contribution has the form (\ref{GR}) with the
replacement $\gamma _{n}\rightarrow i\eta z_{0}$ and omitting the summation
over $n$. As a result, the representation (\ref{GR2}) is valid in the
presence of purely imaginary roots as well.

Similar to the case of the R-region, under the condition
\begin{equation}
\beta /a=1/\left( D/2+\nu \right) ,  \label{SpL}
\end{equation}%
there is a $\lambda =0$ mode with the mode function
\begin{equation}
\varphi _{(\mathrm{L})\sigma }^{(\pm )}(x)=C_{(\mathrm{L})}z^{D/2+\nu
}e^{ik_{r}x^{r}\mp ikt},  \label{phiLsp}
\end{equation}%
where%
\begin{equation}
C_{(\mathrm{L})}^{2}=\frac{(\nu +1)a^{1-D}}{\left( 2\pi \right)
^{p}V_{q}z_{0}^{2\nu +2}k}.  \label{CL2}
\end{equation}%
The contribution of this mode to the Hadamard function is obtained by taking
into account that $\varphi _{(\mathrm{L})\sigma }^{(\pm )}(x)=\Omega _{(%
\mathrm{L})}(z)\varphi _{(\mathrm{M})\sigma }^{(\pm )}(x)$, with%
\begin{equation}
\Omega _{(\mathrm{L})}(z)=\sqrt{2\left( \nu +1\right) }a^{(1-D)/2}\frac{%
z^{D/2+\nu }}{z_{0}^{\nu +1}},  \label{OmL}
\end{equation}%
and is obtained from (\ref{GRs}) by the replacement $\Omega _{(\mathrm{R}%
)}(z)\rightarrow \Omega _{(\mathrm{L})}(z)$. This contribution should be
added to the right-hand side of (\ref{GR}). The representation (\ref{GR2})
is not changed.

\section{Vacuum currents}

\label{sec:Curr}

By using the expressions for the Hadamard function, from formula (\ref{jl1})
we can see that the VEVs of the charge density and of the components of the
current density along non-compact dimensions vanish:%
\begin{equation}
\left\langle j^{l}\right\rangle =0,\;l=0,1,\ldots ,p.  \label{jl00}
\end{equation}
Of course, the latter property for the spatial components is a direct
consequence of the problem symmetry under the reflections $x^{l}\rightarrow
-x^{l}$. For the component along the $l$th compact dimension one finds the
decomposition
\begin{equation}
\langle j^{l}\rangle =\langle j^{l}\rangle _{0}+\langle j^{l}\rangle
_{b},\;l=p+1,\ldots ,D-1,  \label{jldec}
\end{equation}%
where $\langle j^{l}\rangle _{0}$ is the corresponding VEV in the absence of
the brane and the part $\langle j^{l}\rangle _{b}$ is induced by the brane.

The contribution $\langle j^{l}\rangle _{0}$ is investigated in \cite{Beze15}
and is given by the expression%
\begin{eqnarray}
\langle j^{l}\rangle _{0} &=&\frac{4ea^{-1-D}L_{l}}{(2\pi )^{(D+1)/2}}%
\sum_{n_{l}=1}^{\infty }n_{l}\sin (\tilde{\alpha}_{l}n_{l})\sum_{\mathbf{n}%
_{q-1}}\,\cos (\sum_{i\neq l}\tilde{\alpha}_{i}n_{i})  \notag \\
&&\times q_{\nu -1/2}^{(D+1)/2}(1+\sum_{i}n_{i}^{2}L_{i}^{2}/(2z^{2})),
\label{jl0}
\end{eqnarray}%
where $\mathbf{n}_{q-1}=(n_{p+1},\ldots ,n_{l-1},n_{l+1},\ldots ,n_{D-1})$,
and
\begin{equation}
q_{\alpha }^{\mu }(x)=\frac{e^{-i\pi \mu }Q_{\alpha }^{\mu }(x)}{%
(x^{2}-1)^{\mu /2}},  \label{qmu}
\end{equation}%
with $Q_{\alpha }^{\mu }(x)$ being the the associated Legendre function of
the second kind. Near the AdS boundary, $z\rightarrow 0$, the current
density (\ref{jl0}) behaves as $z^{D+2\nu +2}$ and near the horizon the
leading term in the asymptotic expansion is given by $\langle j^{l}\rangle
_{0}\approx (z/a)^{D+1}\langle j^{l}\rangle _{R^{p+1}\times T^{q}}^{(\mathrm{%
M})}$. Here,
\begin{equation}
\langle j^{l}\rangle _{R^{p+1}\times T^{q}}^{(\mathrm{M})}=2eL_{l}\frac{%
\Gamma ((D+1)/2)}{\pi ^{(D+1)/2}}\sum_{n_{l}=1}^{\infty }n_{l}\sin (\tilde{%
\alpha}_{l}n_{l})\sum_{\mathbf{n}_{q-1}}\,\frac{\cos (\sum_{i\neq l}\tilde{%
\alpha}_{i}n_{i})}{(\sum_{i}n_{i}^{2}L_{i}^{2})^{(D+1)/2}},  \label{jlMm0}
\end{equation}%
is the VEV of the current density for a massless scalar field in Minkowski
spacetime with spatial topology $R^{p+1}\times T^{q}$.

Similar to the case of the Hadamard function, we shall consider the
brane-induced contribution in the VEVs of the current density for the R- and
L-regions separately.

\subsection{R-region}

In the R-region the brane-induced contribution in (\ref{jldec}) is obtained
from the corresponding part of the Hadamard function in (\ref{GL3}). By
using the relation
\begin{equation}
\int_{0}^{\infty }d|\mathbf{k}_{p}||\mathbf{k}_{p}|^{p-1}\int_{k}^{\infty }%
\frac{uf(u)du}{\sqrt{u^{2}-k^{2}}}=\frac{\sqrt{\pi }\Gamma \left( p/2\right)
}{2\Gamma \left( (p+1)/2\right) }\int_{0}^{\infty }duu^{p}f(\sqrt{%
u^{2}+k_{(q)}^{2}}),  \label{RelInt}
\end{equation}%
we find the expression%
\begin{equation}
\langle j^{l}\rangle _{b}=-\frac{eC_{p}z^{D+2}}{2^{p-1}a^{D+1}V_{q}}\sum_{%
\mathbf{n}_{q}}k_{l}\int_{k_{(q)}}^{\infty }dx\,x(x^{2}-k_{(q)}^{2})^{\frac{%
p-1}{2}}\frac{\bar{I}_{\nu }(z_{0}x)}{\bar{K}_{\nu }(z_{0}x)}K_{\nu
}^{2}(zx),  \label{jla}
\end{equation}%
with the notation%
\begin{equation}
C_{p}=\frac{\pi ^{-(p+1)/2}}{\Gamma \left( (p+1)/2\right) }.  \label{Cp}
\end{equation}%
From (\ref{jla}) we see that the brane-induced contribution to the current
density along the $l$th compact dimension is an odd periodic function of the
phase $\tilde{\alpha}_{l}$ with the period $2\pi $ and an even periodic
function of the remaining phases $\tilde{\alpha}_{i}$, $i\neq l$, with the
same period. By taking into account the relation (\ref{alfl}), we conclude
that the VEV\ of the current density is a periodic function of the magnetic
flux with the period equal to the flux quantum.

The charge flux through the $(D-1)$-dimensional spatial hypersurface $x^{l}=%
\mathrm{const}$, having the normal $n_{l}=a/z$, is given by the quantity $%
n_{l}\langle j^{l}\rangle $. From (\ref{jla}) it follows that the
corresponding brane-induced contribution, $n_{l}\langle j^{l}\rangle _{b}$,
depends on the lengths of compact dimensions and on the coordinate $z$ in
the form of the ratios $L_{i}/z_{0}$ and $z/z_{0}$. The latter is expressed
in terms of the proper distance from the brane, $y-y_{0}$, as $%
z/z_{0}=e^{(y-y_{0})/a}$.

Let us first consider the flat spacetime limit of (\ref{jla}), corresponding
to the limiting transition $a\rightarrow \infty $ for fixed values of $y$
and $y_{0}$. In this limit, the order of the modified Bessel functions in (%
\ref{jla}) is large. Changing the integration variable to $x\rightarrow \nu
x $, we use the corresponding uniform asymptotic expansions (see, for
instance, \cite{Abra72}). By taking into account that in the limit under
consideration $z\approx a+y$ and $z_{0}\approx a+y_{0}$, to the leading
order we get%
\begin{equation}
\langle j^{l}\rangle _{b}\approx \frac{eC_{p}}{2^{p}V_{q}}\sum_{\mathbf{n}%
_{q}}k_{l}\int_{\sqrt{k_{(q)}^{2}+m^{2}}}^{\infty
}dx\,(x^{2}-k_{(q)}^{2}-m^{2})^{\frac{p-1}{2}}e^{-2x(y-y_{0})}\frac{\beta x+1%
}{\beta x-1}.  \label{MinkLim}
\end{equation}%
The expression in the right-hand side coincides with the boundary-induced
part of the current density for the geometry of a single Robin plate in $%
(D+1)$-dimensional Minkowski spacetime with spatial topology $R^{p+1}\times
T^{q}$ (see \cite{Bell15}).

For a conformally coupled massless field one has $\nu =1/2$ and the modified
Bessel functions in (\ref{jla}) are expressed in terms of the elementary
functions. In this case the expression for the total current density takes
the form%
\begin{equation}
\langle j^{l}\rangle =(z/a)^{D+1}\left[ \langle j^{l}\rangle _{R^{p+1}\times
T^{q}}^{(\mathrm{M})}+\frac{eC_{p}}{2^{p}V_{q}}\sum_{\mathbf{n}%
_{q}}k_{l}\int_{k_{(q)}}^{\infty }dx\,(x^{2}-k_{(q)}^{2})^{\frac{p-1}{2}%
}e^{-2x(z-z_{0})}\frac{\beta _{\mathrm{M}}^{+}x+1}{\beta _{\mathrm{M}}^{+}x-1%
}\right] ,  \label{jlconf}
\end{equation}%
where (the notation with the lower sign is employed below)
\begin{equation}
\beta _{\mathrm{M}}^{\pm }=\frac{\beta z_{0}/a}{1\pm (D-1)\beta /(2a)}.
\label{betM}
\end{equation}%
The right-hand side of (\ref{jlconf}), divided by the conformal factor $%
(z/a)^{D+1}$, coincides with the current density in the corresponding
problem on Minkowski bulk with the plate at $z=z_{0}$ (compare with (\ref%
{MinkLim})) on which the field obeys the Robin boundary condition (\ref{Rob}%
) with the replacement $\beta \rightarrow \beta _{\mathrm{M}}^{+}$. The
difference of the Robin coefficients in the two conformally coupled problems
is related to the fact that this coefficient is not conformally invariant.

At large distances from the brane compared with the AdS curvature radius, $%
y-y_{0}\gg a$, one has $z\gg z_{0}$. In addition, assuming that $z\gg L_{i}$%
, we can see that the dominant contribution to the integral in (\ref{jla})
comes from the region near the lower limit and the contribution of the mode
with a given $\mathbf{n}_{q}$ is suppressed by the factor $e^{-2zk_{(q)}}$.
Under the condition $|\tilde{\alpha}_{i}|<\pi $, assuming that all the
lengths $L_{i}$ are of the same order, the main contribution comes from the
term with $n_{i}=0$, $i=p+1,\ldots ,D-1$, and to the leading order we find%
\begin{equation}
\langle j^{l}\rangle _{b}\approx -\frac{ez^{D-(p-1)/2}\tilde{\alpha}%
_{l}k_{(q)}^{(0)(p-1)/2}}{2^{p+1}\pi ^{(p-1)/2}a^{D+1}V_{q}L_{l}}\frac{\bar{I%
}_{\nu }(z_{0}k_{(q)}^{(0)})}{\bar{K}_{\nu }(z_{0}k_{(q)}^{(0)})}%
e^{-2zk_{(q)}^{(0)}}.  \label{jlbfar}
\end{equation}%
This asymptotic corresponds to points near the AdS horizon. As we have
already mentioned, in this limit, for the boundary-free part one has $%
\langle j^{l}\rangle _{0}\approx (z/a)^{D+1}\langle j^{l}\rangle
_{R^{p+1}\times T^{q}}^{(\mathrm{M})}$, where $\langle j^{l}\rangle
_{R^{p+1}\times T^{q}}^{(\mathrm{M})}$ is the current density for a massless
scalar field in $(D+1)$-dimensional Minkowski spacetime with spatial
topology $R^{p+1}\times T^{q}$ (see (\ref{jlMm0})) and with the lengths of
the compact dimensions $L_{i}$, $i=p+1,\ldots ,D-1$. From here we conclude
that near the horizon the boundary-free part dominates in the total VEV.

For fixed values of $z$ and $L_{i}$, when the location of the brane tends to
the AdS boundary, $z_{0}\rightarrow 0$, to the leading order, from (\ref{jla}%
) one finds%
\begin{eqnarray}
\langle j^{l}\rangle _{b} &\approx &-\frac{4eC_{p}z^{D+2}z_{0}^{2\nu }}{%
2^{2\nu +p}\nu \Gamma ^{2}(\nu )a^{D+1}V_{q}}\frac{A_{0}+\nu B_{0}}{%
A_{0}-\nu B_{0}}\sum_{\mathbf{n}_{q}}k_{l}k_{(q)}^{2\nu +p+1}  \notag \\
&&\times \int_{1}^{\infty }dx\,x^{2\nu +1}(x^{2}-1)^{(p-1)/2}K_{\nu
}^{2}(zk_{(q)}x),  \label{jlAdSbound}
\end{eqnarray}%
and the VEV vanishes as $z_{0}^{2\nu }$.

Now, let us consider the limit when the length of the $l$th dimension is
much smaller than the lengths of the other compact dimensions, $L_{l}\ll
L_{i}$. In this case, in (\ref{jla}) the dominant contribution to the sum
over $n_{i}$, $i=p+1,\ldots ,D-1$, $i\neq l$, comes from large values of $%
|n_{i}|$ and we can replace the summation by the integration in accordance
with
\begin{equation}
\sum_{\mathbf{n}_{q-1}}f(k_{(q-1)})\rightarrow \frac{2^{2-q}\pi
^{-(q-1)/2}V_{q}}{L_{l}\Gamma ((q-1)/2)}\int_{0}^{\infty }du\,u^{q-2}f(u),
\label{SumInt}
\end{equation}%
where $k_{(q-1)}^{2}=k_{(q)}^{2}-k_{l}^{2}$. By making this replacement in (%
\ref{jla}), instead of $x$ we introduce a new integration variable $w$
according to $x=\sqrt{w^{2}+u^{2}+k_{l}^{2}}$. Then, introducing polar
coordinates in the plane $(u,w)$, the integral over the angular variable is
expressed in terms of the gamma function. As a result, to the leading order
we get%
\begin{equation}
\langle j^{l}\rangle _{b}\approx -\frac{eC_{D-2}z^{D+2}}{2^{D-3}a^{D+1}L_{l}}%
\sum_{n_{l}=-\infty }^{+\infty }k_{l}\int_{|k_{l}|}^{\infty
}dx\,x(x^{2}-k_{l}^{2})^{\frac{D-3}{2}}\frac{\bar{I}_{\nu }(z_{0}x)}{\bar{K}%
_{\nu }(z_{0}x)}K_{\nu }^{2}(zx).  \label{Llsmall}
\end{equation}%
The expression in the right-hand side coincides with the brane-induced
contribution in the model with a single compact dimension of the length $%
L_{l}$ ($q=1$, $p=D-2$).

If in addition to $L_{l}\ll L_{i}$ one has $L_{l}\ll z_{0}$, the arguments $%
z_{0}x$ of the modified Bessel functions in (\ref{Llsmall}) are large. By
using the corresponding asymptotic expressions \cite{Abra72}, after the
integration over $x$ we find%
\begin{equation}
\langle j^{l}\rangle _{b}\approx \frac{\left( 1-2\delta _{0B_{0}}\right)
e(z/a)^{D+1}}{2^{D-2}\pi ^{D/2}L_{l}\left( z-z_{0}\right) ^{D/2-1}}%
\sum_{n_{l}=-\infty }^{+\infty }k_{l}|k_{l}|^{D/2-1}K_{D/2-1}(2\left(
z-z_{0}\right) |k_{l}|).  \label{Llsm2}
\end{equation}%
Here, for non-Dirichlet boundary conditions we have assumed that $|\beta
|/a\gg L_{l}/z_{0}$. From (\ref{Llsm2}) it follows that the brane-induced
contribution is located near the brane within the region $z-z_{0}\lesssim
L_{l}$ and has opposite signs for Dirichlet and non-Dirichlet boundary
conditions. At distances $z-z_{0}\gg L_{l}$ it is suppressed by the factor $%
e^{-2(z-z_{0})\tilde{\alpha}_{l}/L_{l}}$. Note that, in the limit $L_{l}\ll
L_{i},z$, for the boundary-free part one has the asymptotic
\begin{equation}
\langle j^{l}\rangle _{0}\approx \frac{2e\Gamma ((D+1)/2)}{\pi
^{(D+1)/2}(a/z)^{D+1}L_{l}^{D}}\sum_{n=1}^{\infty }\frac{\sin (\tilde{\alpha}%
_{l}n)}{n^{D}}.  \label{Llsm0}
\end{equation}

The expression (\ref{jla}) is not convenient for the investigation of the
current density behavior near the brane. To this aim, a more convenient
expression for the VEV of the current density is obtained by using the
representation (\ref{GLalt}) for the Hadamard function. After the
integrations over $w$ and $\mathbf{k}_{p}$, we get the following result%
\begin{eqnarray}
\langle j^{l}\rangle &=&\frac{4ea^{-D-1}z^{D+2}}{(2\pi
)^{p/2+1}V_{q}L_{l}^{p}}\sum_{n=1}^{\infty }\frac{\sin \left( n\tilde{\alpha}%
_{l}\right) }{n^{p+1}}\sum_{\mathbf{n}_{q-1}}\,\int_{0}^{\infty }d\lambda
\,\lambda  \notag \\
&&\times \frac{g_{\nu }^{2}(\lambda z_{0},\lambda z)}{\bar{J}_{\nu
}^{2}(\lambda z_{0})+\bar{Y}_{\nu }^{2}(\lambda z_{0})}g_{p/2+1}(nL_{l}\sqrt{%
\lambda ^{2}+k_{(q-1)}^{2}}),  \label{jlLalt}
\end{eqnarray}%
where we have defined the function%
\begin{equation}
g_{\nu }(x)=x^{\nu }K_{\nu }(x).  \label{genu}
\end{equation}%
In the presence of a bound state, its contribution should be added to (\ref%
{jlLalt}) separately. In the model with a single compact dimension $x^{l}$
of the length $L$ and with the phase $\tilde{\alpha}$, from (\ref{jlLalt})
one finds%
\begin{eqnarray}
\langle j^{l}\rangle &=&\frac{4ea^{-D-1}z^{D+2}}{(2\pi )^{D/2}L^{D-1}}%
\sum_{n=1}^{\infty }\frac{\sin \left( n\tilde{\alpha}\right) }{n^{D-1}}%
\,\int_{0}^{\infty }d\lambda  \notag \\
&&\times \lambda \frac{g_{\nu }^{2}(\lambda z_{0},\lambda
z)g_{D/2}(nL\lambda )}{\bar{J}_{\nu }^{2}(\lambda z_{0})+\bar{Y}_{\nu
}^{2}(\lambda z_{0})},  \label{jl1dim}
\end{eqnarray}%
where we have substituted $p=D-2$.

An important result which follows from (\ref{jlLalt}) is that the VEV of the
current density is finite on the brane. The corresponding value is directly
obtained from (\ref{jlLalt}) putting $z=z_{0}$ and by taking into account
that $g_{\nu }(u,u)=2B_{0}/\pi $. For Dirichlet boundary condition both the
current density and its normal derivative vanish on the brane. The
finiteness of the current density is in clear contrast to the behavior of
the VEVs for the field squared and the energy-momentum tensor which suffer
surface divergences. For example, the VEV of the field squared diverges as $%
1/(z-z_{0})^{D-1}$. The feature that the VEV of the current density is
finite on the brane could be argued on the base of general arguments. In
quantum field theory the ultraviolet divergences in the VEVs of physical
observables bilinear in the field are determined by the local geometrical
characteristics of the bulk and boundary. On the background of standard AdS
geometry with non-compact dimensions the VEV\ of the current density in the
problem under consideration vanishes by the symmetry. The compactification
of the part of spatial dimensions to $q$-dimensional torus does not change
the local bulk and boundary geometries and, consequently, does not add new
divergences to the expectation values compared with the case of trivial
topology.

Under the condition (\ref{betsp}), the contribution of the special mode (\ref%
{ModeSp}) has to be added to the right-hand side of (\ref{jlLalt})
(representation (\ref{jla}) is not changed). This contribution is obtained
from the corresponding part (\ref{GRs}) in the Hadamard function and is
related to the current density for a massless scalar field in $D$%
-dimensional Minkowski spacetime with the spatial topology $R^{p}\times
T^{q} $ by the formula $\langle j^{l}\rangle _{(\mathrm{R})}=\Omega _{(%
\mathrm{R})}^{2}(z)\langle j^{l}\rangle _{R^{p}\times T^{q}}^{(\mathrm{M})}$%
, where the factor $\Omega _{(\mathrm{R})}^{2}(z)$ is given by the
expression (\ref{Omz}). By using the expression for $\langle j^{l}\rangle
_{R^{p}\times T^{q}}^{(\mathrm{M})}$ from \cite{Beze13c}, one gets%
\begin{equation}
\langle j^{l}\rangle _{(\mathrm{R})}=\frac{8e\left( \nu -1\right)
z_{0}^{2\nu -2}z^{D-2\nu }}{(2\pi )^{(p+3)/2}a^{D-1}V_{q}L_{l}^{p}}%
\sum_{n=1}^{\infty }\frac{\sin \left( n\tilde{\alpha}_{l}\right) }{n^{p+1}}%
\sum_{\mathbf{n}_{q-1}}g_{p/2+1}(nL_{l}k_{(q-1)}).  \label{jls}
\end{equation}%
Recall that, as the necessary condition for the presence of this
contribution we have $\nu >1$. In the model with a single compact dimension
this gives%
\begin{equation}
\langle j^{l}\rangle _{(\mathrm{R})}=\frac{4e\left( \nu -1\right) \Gamma
(D/2)}{\pi ^{D/2}a^{D-1}L^{D-1}}z_{0}^{2\nu -2}z^{D-2\nu }\sum_{n=1}^{\infty
}\frac{\sin \left( n\tilde{\alpha}_{l}\right) }{n^{D-1}}.  \label{jls1}
\end{equation}%
Note that in the case $\nu =D/2$, the current density from the special mode
does not depend on $z$.

Let us consider the behavior of the current density in the limit when both
the location of the brane and the point of observation are close to the AdS
boundary, $z_{0},z\ll L_{i}$. Under this condition, the arguments of the
Bessel functions $J_{\nu }(x)$ and $Y_{\nu }(x)$ in (\ref{jlLalt}) are small
and we use the corresponding asymptotic expressions. To the leading order,
the integral over $\lambda $ is expressed in terms of the Macdonald function
and we get the expression%
\begin{eqnarray}
\langle j^{l}\rangle &\approx &\frac{ea^{-D-1}z^{D-2\nu +2}L_{l}^{-p-2\nu -2}%
}{2^{p/2+\nu -1}\pi ^{p/2+1}\Gamma (\nu +1)V_{q}}\left( z^{2\nu }-\frac{%
A_{0}+\nu B_{0}}{A_{0}-\nu B_{0}}z_{0}^{2\nu }\right) ^{2}  \notag \\
&&\times \sum_{n=1}^{\infty }\frac{\sin \left( n\tilde{\alpha}_{l}\right) }{%
n^{p+2\nu +3}}\sum_{\mathbf{n}_{q-1}}g_{p/2+\nu +2}(nL_{l}k_{(q-1)}).
\label{zz0sm}
\end{eqnarray}%
The part with the term $z^{2\nu }$ in the brackets gives the asymptotic for
the VEV in the absence of the brane.

The representation (\ref{jlLalt}) is also well-suited for the investigation
of the asymptotic behavior in the limit of large $L_{l}$ compared with the
other length scales of the model. In this limit the argument of the function
$g_{\mu }(x)$ in (\ref{jlLalt}) is large and we can use the asymptotic
expression $g_{\mu }(x)\approx \sqrt{\pi /2}x^{\mu -1/2}e^{-x}$, for $x\gg 1$%
. The dominant contribution to the integral comes from the region near the
lower limit. Let us denote by $k_{(q-1)}^{(0)}$ the lowest value of $%
k_{(q-1)}$, $k_{(q-1)}^{(0)}=\mathrm{min}(k_{(q-1)})$. For $|\tilde{\alpha}%
_{i}|<\pi $ this value is realized by the mode with $n_{i}=0$, $i\neq l$,
and we have%
\begin{equation}
k_{(q-1)}^{(0)2}=\sum_{i=p+1,\neq l}^{D-1}\tilde{\alpha}_{i}^{2}/L_{i}^{2}.
\label{kq-10}
\end{equation}%
Two cases should be considered separately. For $k_{(q-1)}^{(0)}\neq 0$, in
the series over $n$ the leading contribution comes from the $n=1$ term and
we get%
\begin{eqnarray}
\langle j^{l}\rangle &\approx &\frac{2ea^{-D-1}z^{D-2\nu
+2}k_{(q-1)}^{(0)(p+3)/2+\nu }}{\pi ^{(p+1)/2}\Gamma (\nu
+1)V_{q}(2L_{l})^{(p+1)/2+\nu }}  \notag \\
&&\times \left( z^{2\nu }-\frac{A_{0}+\nu B_{0}}{A_{0}-\nu B_{0}}z_{0}^{2\nu
}\right) ^{2}\frac{\sin \tilde{\alpha}_{l}}{e^{L_{l}k_{(q-1)}^{(0)}}}.
\label{jlLlarge}
\end{eqnarray}%
In this case the current density is exponentially small. Note that (\ref%
{jlLlarge}) is also obtained from (\ref{zz0sm}) in the limit of large $L_{l}$%
. For $k_{(q-1)}^{(0)}=0$, by using the standard integral involving the
Macdonald function \cite{Prud86}, for the leading term in the asymptotic
expansion one finds the expression%
\begin{eqnarray}
\langle j^{l}\rangle &\approx &\frac{4e\Gamma \left( p/2+\nu +2\right)
a^{-D-1}z^{D-2\nu +2}}{\pi ^{p/2+1}\Gamma (\nu +1)V_{q}L_{l}^{p+2\nu +2}}
\notag \\
&&\times \left( z^{2\nu }-\frac{A_{0}+\nu B_{0}}{A_{0}-\nu B_{0}}z_{0}^{2\nu
}\right) ^{2}\sum_{n=1}^{\infty }\frac{\sin \left( n\tilde{\alpha}%
_{l}\right) }{n^{p+2\nu +3}}\,.  \label{jlLlarge2}
\end{eqnarray}%
Now the decay of the current density with increasing $L_{l}$ is power law
for both massless and massive fields. This result for massive fields is in
contrast to the corresponding behavior of the current density in Minkowski
bulk. In the latter geometry the decay of the current density is exponential
with the factor $e^{-mL_{l}}$.

For points outside the brane, $z>z_{0}$, another expression for the current
density is obtained from the representation (\ref{Gl}). After evaluating the
integrals, one finds the following result%
\begin{eqnarray}
\langle j^{l}\rangle &=&\langle j^{l}\rangle _{0}-\frac{%
4ea^{-1-D}L_{l}^{-p}z^{D+2}}{(2\pi )^{p/2+1}V_{q}}\sum_{n=1}^{\infty }\frac{%
\sin \left( n\tilde{\alpha}_{l}\right) }{n^{p+1}}\sum_{\mathbf{n}%
_{q-1}}\int_{k_{(q-1)}}^{\infty }dx\,  \notag \\
&&\times x\frac{\bar{I}_{\nu }(xz_{0})}{\bar{K}_{\nu }(xz_{0})}K_{\nu
}^{2}(xz)w_{p/2+1}(nL_{l}\sqrt{x^{2}-k_{(q-1)}^{2}}).  \label{jLalt2}
\end{eqnarray}%
with the function
\begin{equation}
w_{\nu }(x)=x^{\nu }J_{\nu }(x).  \label{w}
\end{equation}%
In the absence of the bound states, the equivalence of the representations (%
\ref{jla}) and (\ref{jLalt2}) for the brane-induced contribution is directly
seen by using the formula%
\begin{equation}
\sum_{n_{l}=-\infty }^{+\infty }k_{l}g(|k_{l}|)=\frac{2L_{l}}{\pi }%
\sum_{n=1}^{\infty }\sin (n\tilde{\alpha}_{l})\int_{0}^{\infty }dx\,x\sin
(nL_{l}x)g(x).  \label{SumForm3}
\end{equation}%
The latter relation follows from the Poisson's resummation formula (see also
\cite{Bell15}).

In what follows, all the graphs are plotted in the $D=4$ model with a single
compact dimension of the coordinate length $L$ and with the phase $\tilde{%
\alpha}$, for a minimally coupled ($\xi =0$) massless scalar field. For the
corresponding value of the parameter $\nu $ one has $\nu =D/2=2$.

\begin{figure}[tbph]
\begin{center}
\begin{tabular}{cc}
\epsfig{figure=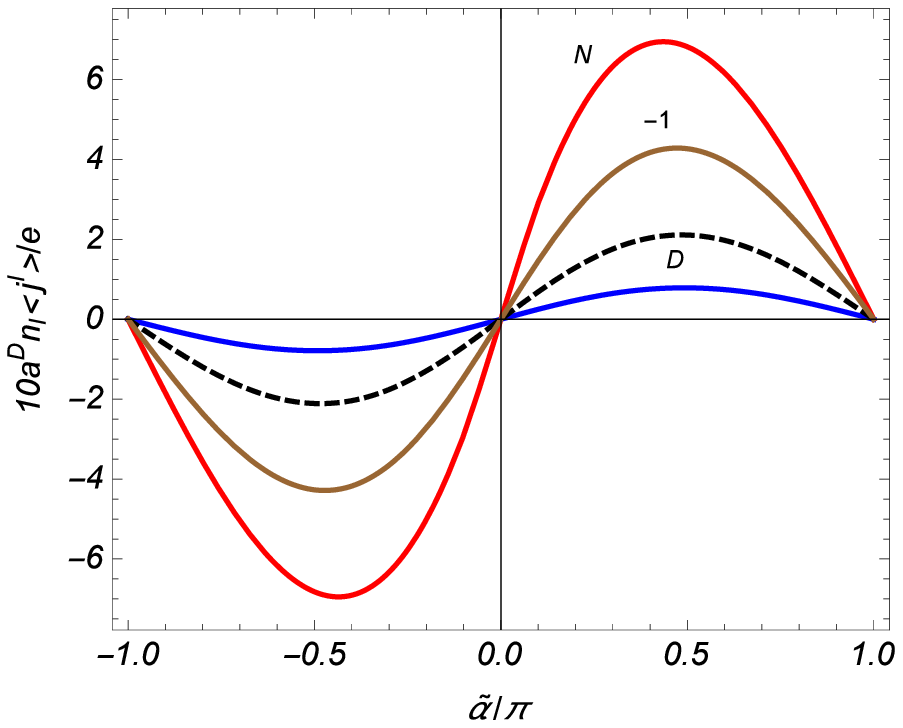,width=7.cm,height=5.5cm} & \quad %
\epsfig{figure=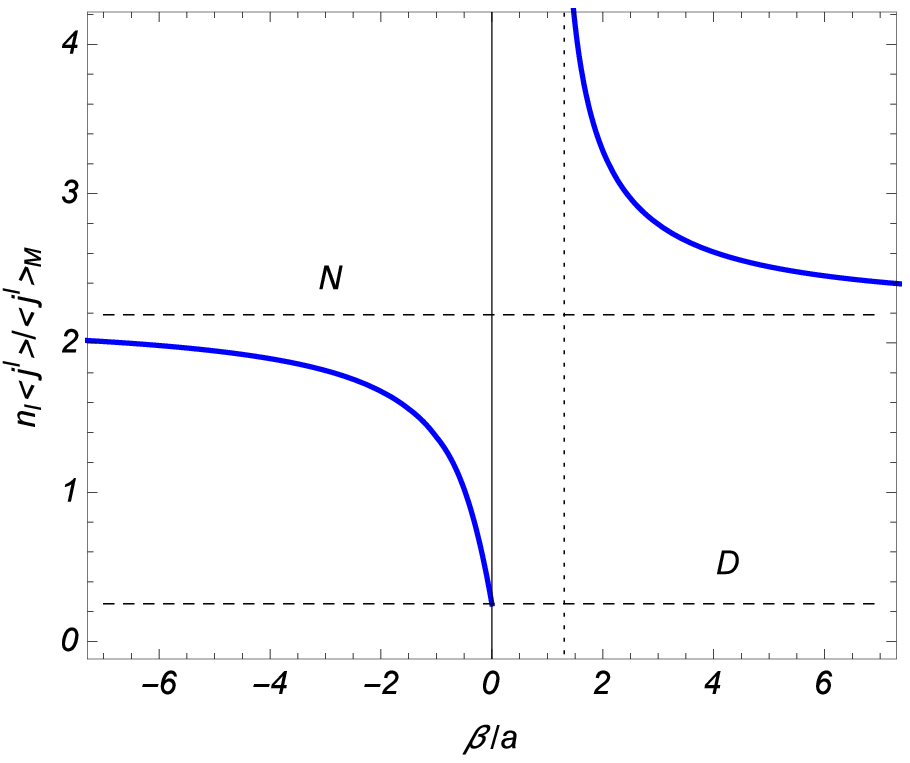,width=7.cm,height=5.5cm}%
\end{tabular}%
\end{center}
\caption{The VEV of the current density as a function of the phase in the
periodicity condition (left panel) for $D=4$ AdS space with a single compact
dimension and for Dirichlet, Neumann and Robin ($\protect\beta /a=-1$)
boundary conditions. The graphs are plotted for $z_{0}/L=1$, $z/z_{0}=1.2$.
The right panel displays the ratio $n_{l}\langle j^{l}\rangle /\langle
j^{l}\rangle _{\mathrm{M}}$ as a function of the coefficient in Robin
boundary condition for fixed values $\tilde{\protect\alpha}=\protect\pi /2$,
$z/z_{0}=1.2$, $z_{0}/L=1$.}
\label{fig2}
\end{figure}

On the left panel of figure \ref{fig2} we have depicted the current density
as a function of the phase $\tilde{\alpha}$ for fixed values of the
parameters $z_{0}/L=1$, $z/z_{0}=1.2$. The graphs are plotted for Dirichlet
(D), Neumann (N) and for Robin (with $\beta /a=-1$, the number near the
curve) boundary conditions. The dashed curve presents the current density in
the same model when the brane is absent. As is seen, depending on the
boundary condition, the presence of the brane leads to the increase or
decrease of the current density. Note that for the example considered one
has $\nu -D/2=0$ and, in accordance with (\ref{betsp}), for Neumann boundary
condition there is a special mode (\ref{ModeSp}) the contribution of which
is given by (\ref{jls}). This contribution should be added to (\ref{jl1dim}%
). Alternatively, in the numerical evaluation we may use the representation (%
\ref{jla}) which holds in the presence of the special mode as well. We have
numerically checked that both these ways of the evaluation give the same
result.

The right panel of figure \ref{fig2} presents the ratio $n_{l}\langle
j^{l}\rangle /\langle j^{l}\rangle _{\mathrm{M}}$ as a function of the
coefficient in Robin boundary condition, measured in units of AdS curvature
radius. Here,
\begin{equation}
\langle j^{l}\rangle _{\mathrm{M}}=\frac{2e\Gamma ((D+1)/2)}{\pi
^{(D+1)/2}(aL/z)^{D}}\sum_{n=1}^{\infty }\frac{\sin (\tilde{\alpha}n)}{n^{D}}%
,  \label{jlM}
\end{equation}%
is the current density for a massless scalar field in $(D+1)$-dimensional
Minkowski spacetime with topology $R^{D-1}\times S^{1}$ and with the length
of the compact dimension $aL/z$. Note that the latter is the proper length
of the compact dimension in AdS spacetime measured by an observer with a
given $z$. The graph is plotted for fixed values $\tilde{\alpha}=\pi /2$, $%
z/z_{0}=1.2$, $z_{0}/L=1$. The vertical dotted curve corresponds to the
critical value $\beta _{\mathrm{R}}/a\approx 1.31$. In the region $0<\beta
<\beta _{\mathrm{R}}$ the vacuum is unstable. The horizontal dashed curves
correspond to Dirichlet and Neumann boundary conditions. As expected, in the
limits $\beta \rightarrow 0$ and $\beta \rightarrow \infty $ the results for
Dirichlet and Neumann conditions are obtained. For Dirichlet boundary
condition the current density takes its minimal value (the minimal absolute
value for negative $\tilde{\alpha}$).

\begin{figure}[tbph]
\begin{center}
\begin{tabular}{cc}
\epsfig{figure=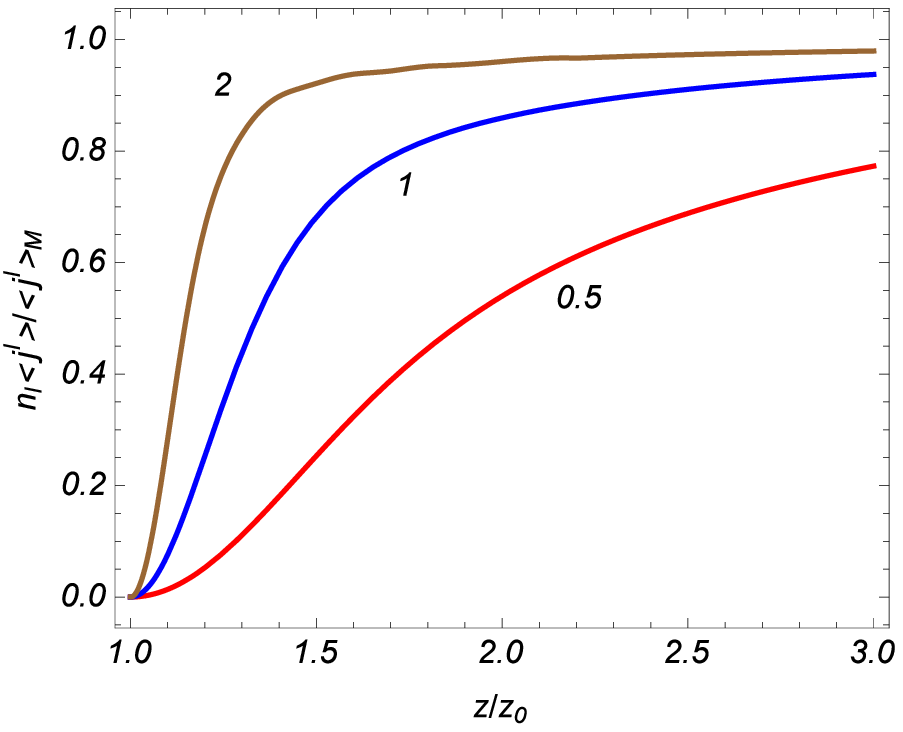,width=7.cm,height=5.5cm} & \quad %
\epsfig{figure=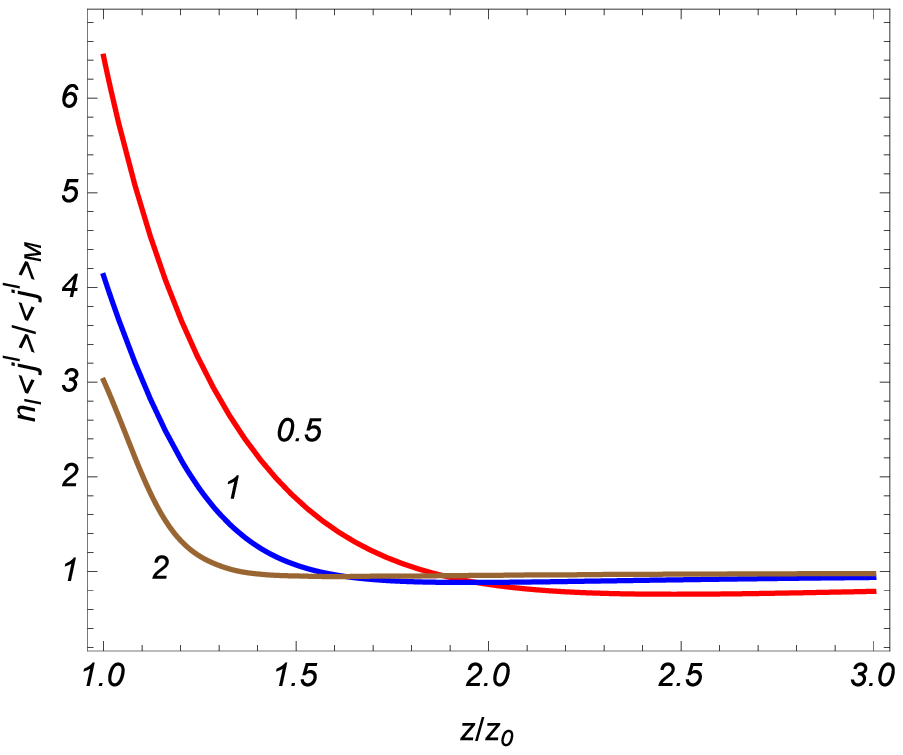,width=7.cm,height=5.5cm}%
\end{tabular}%
\end{center}
\caption{The dependence of the quantity $n_{l}\langle j^{l}\rangle /\langle
j^{l}\rangle _{\mathrm{M}}$ on $z/z_{0}$ for Dirichlet (left panel) and
Neumann (right panel) boundary conditions. The graphs are plotted for fixed
values of $z_{0}/L$ (figures near the curves) and for $\tilde{\protect\alpha}%
=\protect\pi /2$.}
\label{fig3}
\end{figure}

In figure \ref{fig3} we have plotted the ratio $n_{l}\langle j^{l}\rangle
/\langle j^{l}\rangle _{\mathrm{M}}$ as a function of $z/z_{0}$ for
Dirichlet (left panel) and Neumann (right panel) boundary conditions on the
brane. The numbers near the curves are the values of $z_{0}/L$. For the
phase we have taken the value $\tilde{\alpha}=\pi /2$. As it has been
already mentioned, for Dirichlet boundary condition the current density
vanishes on the brane. From the asymptotic analysis given above it follows
that on the horizon, $z/z_{0}\rightarrow \infty $, one has $n_{l}\langle
j^{l}\rangle /\langle j^{l}\rangle _{\mathrm{M}}\rightarrow 1$. This
behavior is seen from the graphs.

\begin{figure}[tbph]
\begin{center}
\epsfig{figure=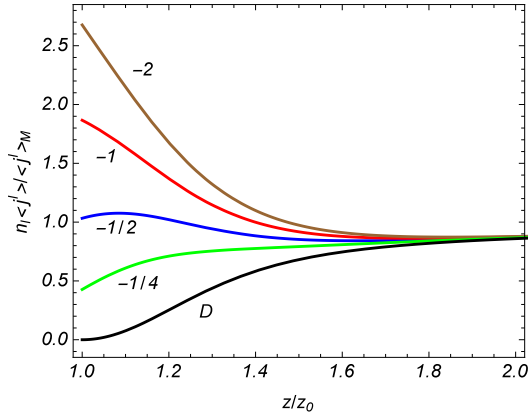,width=7.cm,height=5.5cm}
\end{center}
\caption{The same as in figure \protect\ref{fig3} for Robin boundary
condition. The graphs are plotted for $z_{0}/L=1$ and the numbers near the
curves correspond to the values of $\protect\beta /a$.}
\label{fig4}
\end{figure}

Figure \ref{fig4} displays the dependence of $n_{l}\langle j^{l}\rangle
/\langle j^{l}\rangle _{\mathrm{M}}$ on $z/z_{0}$ in the case of Robin
boundary condition for several values of $\beta /a$ (numbers near the
curves). For $\beta /a=-2/(D-1)$ the normal derivative $\partial _{z}\left(
n_{l}\langle j^{l}\rangle /\langle j^{l}\rangle _{\mathrm{M}}\right) $
vanishes on the brane. This can also be directly seen from the analytic
expression by taking into account that $v\partial _{v}g_{\nu
}(u,v)|_{v=u}=-2A_{0}/\pi $.

\begin{figure}[tbph]
\begin{center}
\epsfig{figure=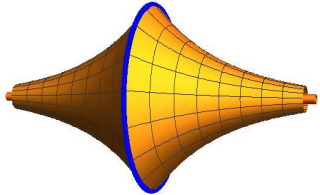,width=5.cm,height=4cm}
\end{center}
\caption{$D=2$ spatial geometry corresponding to the Randall--Sundrum
1-brane model with a compactified dimension.}
\label{fig5}
\end{figure}

From the results derived in this section we can obtain the current density
in $Z_{2}$-symmetric braneworld models of the Randall--Sundrum type with a
single brane. In the original Randall--Sundrum 1-brane model \cite{Rand99}
the universe is realized as a $Z_{2}$-symmetric positive tension brane in
5-dimensional AdS spacetime. In this simplest variant the only contribution
to the curvature comes from the negative cosmological constant in the bulk.
However, most scenarios motivated from string theories predict the presence
of other bulk fields, such as scalar fields. In addition, they predict also
small compact dimensions originating from 10D string backgrounds. In a
generalized $(D+1)$-dimensional version of the Randall--Sundrum 1-brane
model the line element is given by (\ref{metric}) with the warp factor $%
e^{-|y-y_{0}|/a}$ where $y_{0}$ is the location of the brane. The background
geometry contains two patches $y>y_{0}$ of the AdS glued by the brane and
related by the $Z_{2}$-symmetry identification $y-y_{0}\longleftrightarrow
y_{0}-y$. The corresponding spatial geometry in the case $D=2$, embedded
into a 3D Euclidean space is depicted in figure \ref{fig5}.

Because of the absolute value sign in the exponent of the warp factor, the
Ricci scalar contains a contribution located on the brane,%
\begin{equation}
R=4D\delta (y-y_{0})/a-D(D+1)/a^{2}.  \label{RRS}
\end{equation}%
For non-minimally coupled scalar fields, this leads to delta-type terms in
the field equation. An additional delta-type term may come from the boundary
action of a scalar field of the form $S_{b}=c\int d^{D}xdy\,\sqrt{|g|}%
\varphi ^{2}\delta (y-y_{0})$, where $c$ is the so-called brane mass term.
The boundary condition for the mode functions is obtained by integrating the
field equation near the brane. In a way similar to that used in \cite%
{Gher00,Flac01,Saha05}, it can be seen that for fields even under the
reflection with respect to the brane (untwisted scalar field) the boundary
condition is of the Robin type with%
\begin{equation}
\beta =-1/(c+2D\xi /a).  \label{betRS}
\end{equation}%
In particular, for a minimally coupled field with the zero brane mass term
the boundary condition is the Neumann one. For fields odd with respect to
the reflection (twisted fields) the boundary condition is reduced to the
Dirichlet one. Now, in the $Z_{2}$-symmetric model the integration over $y$
in the normalization integral (\ref{Norm}) goes over the interval $(-\infty
,+\infty )$. As a result the square of the normalization coefficient
contains an additional factor $1/2$ compared to the one we have obtained for
the R-region. Hence, the expressions for the VEV\ of the current density in
the generalized Randall--Sundrum 1-brane model with compact dimensions are
obtained from those given above in this section with an additional factor $%
1/2$ and with the Robin coefficient (\ref{betRS}) for untwisted fields and
with $\beta =0$ for twisted fields.

\subsection{L-region}

Now we turn to the current density in the L-region. By using the expression (%
\ref{GR2}) for the Hadamard function, the current density in this region is
decomposed as (\ref{jldec}) with the brane-induced part%
\begin{equation}
\langle j^{l}\rangle _{b}=-\frac{eC_{p}z^{D+2}}{2^{p-1}a^{D+1}V_{q}}\sum_{%
\mathbf{n}_{q}}k_{l}\int_{k_{(q)}}^{\infty }dx\,x(x^{2}-k_{(q)}^{2})^{\frac{%
p-1}{2}}\frac{\bar{K}_{\nu }(z_{0}x)}{\bar{I}_{\nu }(z_{0}x)}I_{\nu
}^{2}(zx).  \label{jlb1}
\end{equation}%
Under the condition $\eta <k_{(q)}^{(0)}$, this representation is valid in
the presence of the mode with $\lambda =i\eta $, where $\eta $ is the zero
of the function $\bar{I}_{\nu }(z_{0}\eta )$, and also in the presence of
the special mode (\ref{phiLsp}). For large values of AdS radius $a$, in a
way similar to that for the R-region, we can see the limiting transition of
the expression (\ref{jlb1}) to the corresponding formula for a plate in
Minkowski bulk.

For a conformally coupled massless field, the expression of the total
current density takes the form%
\begin{eqnarray}
\langle j^{l}\rangle &=&(z/a)^{D+1}\bigg\{\langle j^{l}\rangle
_{R^{p+1}\times T^{q}}^{(\mathrm{M})}-\frac{eC_{p}}{2^{p}V_{q}}\sum_{\mathbf{%
n}_{q}}k_{l}\int_{k_{(q)}}^{\infty }dx\,  \notag \\
&&\times (x^{2}-k_{(q)}^{2})^{\frac{p-1}{2}}\bigg[e^{-2zx}+\frac{4\sinh
^{2}(zx)}{\frac{1-\beta _{\mathrm{M}}^{-}x}{1+\beta _{\mathrm{M}}^{-}x}%
e^{2z_{0}x}-1}\bigg]\bigg\},  \label{jlconf2}
\end{eqnarray}%
with $\beta _{\mathrm{M}}^{-}$ defined by (\ref{betM}). Here, the first term
in the figure braces and the part with the first term in the square brackets
come from $\langle j^{l}\rangle _{0}$. The expression on the right of (\ref%
{jlconf2}), divided by the conformal factor $(z/a)^{D+1}$, coincides with
the current density in the region between two plates on Minkowski bulk with
Dirichlet boundary condition on the left plate and Robin condition (\ref{Rob}%
), with $\beta \rightarrow \beta _{\mathrm{M}}^{-}$, on the right one (see
\cite{Bell15} for the problem with Robin boundary conditions on both
plates). The fact that the problem with a single brane in AdS bulk in the
L-region is conformaly related to the problem with two plates in Minkowski
bulk is a consequence of the boundary condition we have imposed on the AdS
boundary.

The asymptotic behavior of the VEV near the AdS boundary, $z\rightarrow 0$,
is directly obtained from (\ref{jlb1}), by using the expression of the
modified Bessel function for small arguments. To the leading order we get%
\begin{eqnarray}
\langle j^{l}\rangle _{b} &\approx &-\frac{2^{1-2\nu -p}eC_{p}z^{D+2\nu +2}}{%
a^{D+1}V_{q}\Gamma ^{2}(\nu +1)}\sum_{\mathbf{n}_{q}}k_{l}k_{(q)}^{2\nu +p+1}
\notag \\
&&\times \int_{1}^{\infty }dx\,x^{2\nu +1}(x^{2}-1)^{(p-1)/2}\frac{\bar{K}%
_{\nu }(z_{0}k_{(q)}x)}{\bar{I}_{\nu }(z_{0}k_{(q)}x)},  \label{jlbAdSbound}
\end{eqnarray}%
and the brane-induced contribution vanishes as $z^{D+2\nu +2}$. Recall that
near the AdS boundary the part $\langle j^{l}\rangle _{0}$ in the VEV of the
current density behaves in a similar way and, hence, on the AdS boundary the
ratio of the brane-induced and boundary-free contributions tend to a finite
limiting value.

In the limit when the brane tends to the AdS horizon, $z_{0}\rightarrow
\infty $, the argument $z_{0}x$ of the modified Bessel functions is large.
By using the corresponding asymptotics and by taking into account that the
dominant contribution to the integral in (\ref{jlb1}) comes from the region
near the lower limit, we can see that the leading order term is given by%
\begin{equation}
\langle j^{l}\rangle _{b}\approx \frac{(1-2\delta _{0B_{0}})e\tilde{\alpha}%
_{l}z^{D+2}e^{-2z_{0}k_{(q)}^{(0)}}}{2^{p}\pi
^{(p-1)/2}a^{D+1}V_{q}L_{l}z_{0}^{(p+1)/2}}k_{(q)}^{(0)(p+1)/2}I_{\nu
}^{2}(zk_{(q)}^{(0)}).  \label{jlbNearHor}
\end{equation}%
Hence, for a fixed value of $z$, when the brane location tends to the AdS
horizon, the brane-induced contribution is exponentially suppressed.

If the length of the $l$th dimension is much smaller than the lengths of the
remaining compact dimensions, $L_{l}\ll L_{i}$, in a way similar to that for
the R-region, we can see that, to the leading order, the brane-induced
contribution coincides with the corresponding quantity in the model with a
single compact dimension of the length $L_{l}$. The expression for the
latter is obtained from the right-hand side of (\ref{Llsmall}) by the
replacements $I_{\nu }\rightleftarrows K_{\nu }$. If in addition $L_{l}\ll z$%
, the corresponding asymptotic expression is given by the right-hand side of
(\ref{Llsm2}) with $z_{0}-z$ instead of $z-z_{0}$, and the brane-induced
contribution is concentrated near the brane in the region $z_{0}-z\lesssim
L_{l}$.

The representation (\ref{jlb1}) is not well adapted for the investigation of
the asymptotic near the brane. A more suitable representation is obtained by
using the formula (\ref{GRalt}) for the Hadamard function:%
\begin{eqnarray}
\langle j^{l}\rangle &=&\frac{16ea^{-1-D}z^{D+2}}{\left( 2\pi \right)
^{p/2+1}V_{q}L_{l}^{p}z_{0}^{2}}\sum_{n=1}^{\infty }\frac{\sin \left( n%
\tilde{\alpha}_{l}\right) }{n^{p+1}}\sum_{\mathbf{n}_{q-1}}\sum_{i}\,\gamma
_{i}  \notag \\
&&\times T_{\nu }(\gamma _{i})J_{\nu }^{2}(\gamma
_{i}z/z_{0})g_{p/2+1}(nL_{l}\sqrt{\gamma _{i}^{2}/z_{0}^{2}+k_{(q-1)}^{2}}).
\label{jlbalt}
\end{eqnarray}%
Unlike to the representation (\ref{jlb1}), in the presence of a bound state,
its contribution must be additionally added to (\ref{jlbalt}). The latter is
obtained from the right-hand side of (\ref{jlbalt}) by the replacement $%
\gamma _{i}\rightarrow i\eta z_{0}$ and omitting the summation over $i$. The
corresponding representation is valid under the condition $\eta
<k_{(q-1)}^{(0)}$. From (\ref{jlbalt}) we conclude that the VEV of the
current density is finite on the brane. Similar to the VEV in the R-region,
the current density and its normal derivative vanish on the brane for
Dirichlet boundary condition. Another representation is obtained from (\ref%
{jLalt2}) by the replacements $I_{\nu }\rightleftarrows K_{\nu }$ and with
the replacement $\beta \rightarrow -\beta $ in the notations with overbars..

For the value of the Robin coefficient (\ref{SpL}) there is a special mode (%
\ref{phiLsp}). As we have seen above, the corresponding contribution to the
Hadamard function is expressed in terms of the Hadamard function for a
massless scalar field in $D$-dimensional Minkowski spacetime with the
spatial topology $R^{p}\times T^{q}$. By using this relation, for the
contribution of the special mode to the current density in the L-region we
find%
\begin{equation}
\langle j^{l}\rangle _{(\mathrm{L})}=\frac{8e\left( \nu +1\right)
z_{0}^{-2\nu -2}z^{D+2\nu }}{(2\pi )^{(p+3)/2}a^{D-1}V_{q}L_{l}^{p}}%
\sum_{n=1}^{\infty }\frac{\sin \left( n\tilde{\alpha}_{l}\right) }{n^{p+1}}%
\sum_{\mathbf{n}_{q-1}}g_{p/2+1}(nL_{l}k_{(q-1)}).  \label{jlLs}
\end{equation}%
In the case of a single compact dimension this simplifies to%
\begin{equation}
\langle j^{l}\rangle _{(\mathrm{L})}=\frac{4e\left( \nu +1\right) \Gamma
(D/2)z^{D+2\nu }}{\pi ^{D/2}a^{D-1}L^{D-1}z_{0}^{2\nu +2}}\sum_{n=1}^{\infty
}\frac{\sin \left( n\tilde{\alpha}_{l}\right) }{n^{D-1}}.  \label{jLs1}
\end{equation}%
Under the condition (\ref{SpL}), the contribution (\ref{jlLs}) should be
added to the right-hand side of (\ref{jlbalt}).

For the investigation of the asymptotic behavior for the contribution of the
modes with $\lambda =\gamma _{n}/a$ in the limit of large values of $L_{l}$,
compared with the other length scales, it is convenient to use the
representation (\ref{jlbalt}). By using the asymptotic expression of the
function $g_{\mu }(x)$ for large values of the argument, we can see that the
dominant contribution comes from the lowest mode with $i=1$ and to the
leading order one finds%
\begin{equation}
\langle j^{l}\rangle =\frac{8ez^{D+2}\gamma _{1}T_{\nu }(\gamma _{1})\sin
\tilde{\alpha}_{l}}{\left( 2\pi \right)
^{(p+1)/2}V_{q}L_{l}^{p}z_{0}^{2}a^{D+1}}\,\frac{x^{(p+1)/2}}{e^{-x}}J_{\nu
}^{2}(\gamma _{1}z/z_{0}),  \label{jlbLlarge}
\end{equation}%
with $x=L_{l}\sqrt{\gamma _{1}^{2}/z_{0}^{2}+k_{(q-1)}^{(0)2}}$. Hence,
unlike to the R-region, the decay of the current density is exponential for
both cases $k_{(q-1)}^{(0)}=0$ and $k_{(q-1)}^{(0)}\neq 0$. Under the
condition (\ref{SpL}) one has the additional contribution (\ref{jlLs}) from
the mode with $\lambda =0$. For $k_{(q-1)}^{(0)}\neq 0$, in the limit of
large $L_{l}$, this contribution decays exponentially, as $%
e^{-L_{l}k_{(q-1)}^{(0)}}$. In the case $k_{(q-1)}^{(0)}=0$ the decay is
power law, like $1/L_{l}^{p+1}$. In both cases the contribution of the
special mode dominates in the total VEV.

Now let us consider the asymptotic of the current density (\ref{jlbalt})
when the brane is close to AdS boundary, $z_{0}\ll L_{i}$. The dominant
contribution comes from large values of $|n_{j}|$, $j\neq l$, and we can
replace the summation over $\mathbf{n}_{q-1}$ by the integration in
accordance with (\ref{SumInt}). After the integration over $k_{(q-1)}$ we get%
\begin{equation}
\langle j^{l}\rangle \approx \frac{16ea^{-1-D}z^{D+2}}{(2\pi
)^{D/2}L_{l}^{D-1}z_{0}^{2}}\sum_{n=1}^{\infty }\frac{\sin \left( n\tilde{%
\alpha}_{l}\right) }{n^{D-1}}\sum_{i}\,\gamma _{i}T_{\nu }(\gamma
_{i})J_{\nu }^{2}(\gamma _{i}z/z_{0})g_{D/2}(nL_{l}\gamma _{i}/z_{0}).
\label{jlBhor}
\end{equation}%
The right-hand side presents the brane-induced contribution in the model
with a single compact dimension of the length $L_{l}$. In the limit under
consideration the argument of the function $g_{D/2}(x)$ is large. By using
the corresponding asymptotic expression, we see that the main contribution
comes from the term with $n=1$, $i=1$ with the result%
\begin{equation}
\langle j^{l}\rangle \approx \frac{8ea^{-1-D}z^{D+2}\sin \tilde{\alpha}_{l}}{%
(2\pi )^{(D-1)/2}L_{l}^{D+1}}\,\frac{(\gamma _{1}L_{l}/z_{0})^{(D+3)/2}}{%
\gamma _{1}e^{\gamma _{1}L_{l}/z_{0}}}T_{\nu }(\gamma _{1})J_{\nu
}^{2}(\gamma _{1}z/z_{0}).  \label{jlBhor2}
\end{equation}%
For the Robin boundary condition with (\ref{SpL}), the asymptotic for the
contribution from the special mode is directly obtained from (\ref{jlLs}).
This contribution behaves as $(z_{0}/L_{l})^{D-2}(z/z_{0})^{D+2\nu }$ and
the corresponding decay, as a function of $z_{0}$ (for a fixed $z/z_{0}$) is
power law.

\begin{figure}[tbph]
\begin{center}
\begin{tabular}{cc}
\epsfig{figure=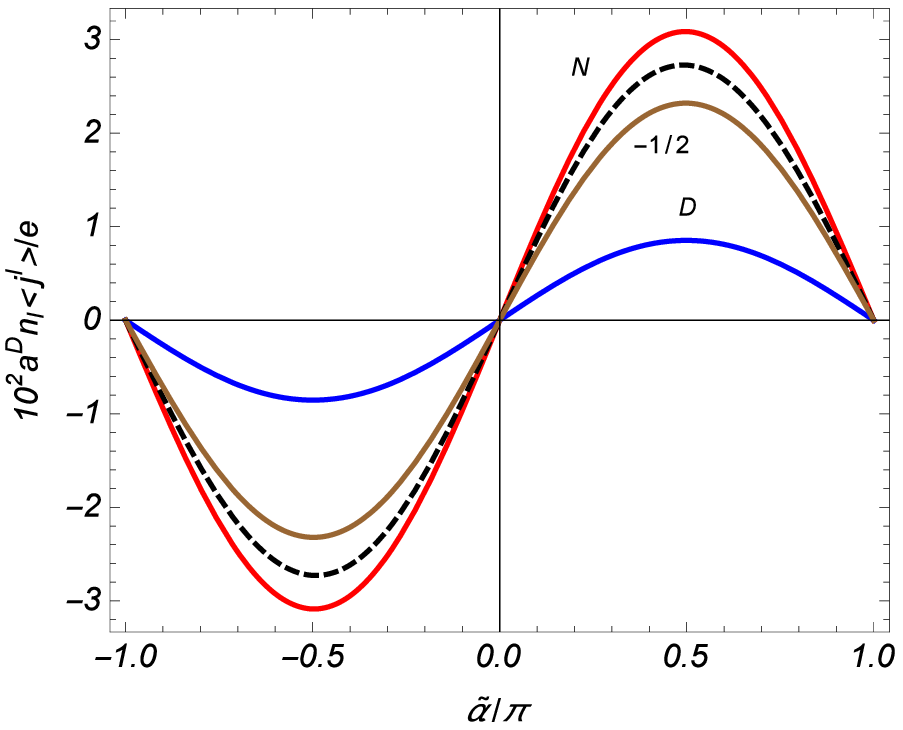,width=7.cm,height=5.5cm} & \quad %
\epsfig{figure=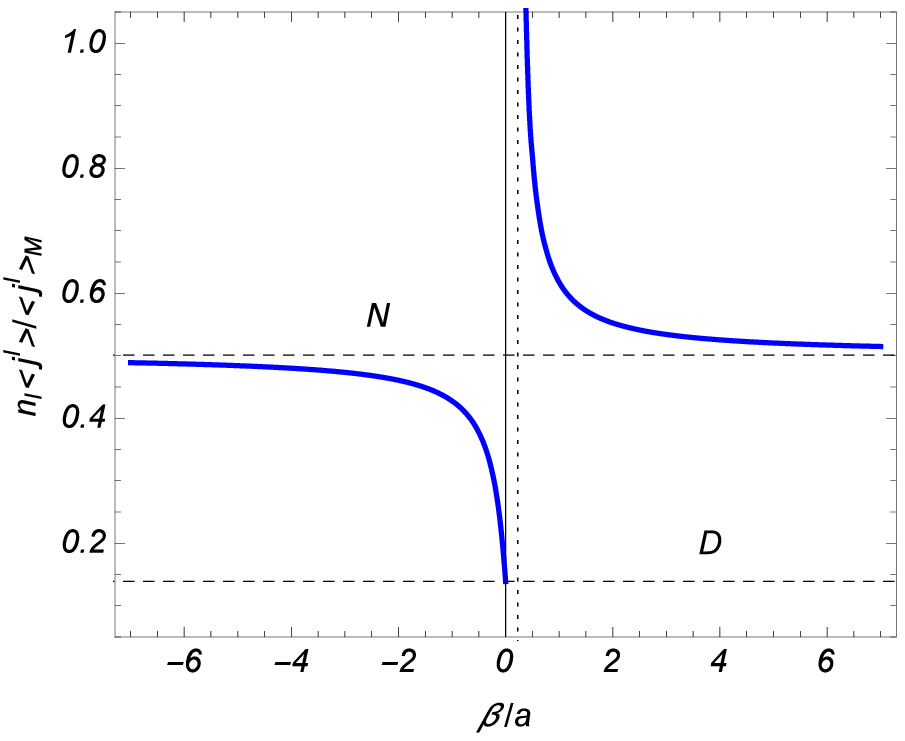,width=7.cm,height=5.5cm}%
\end{tabular}%
\end{center}
\caption{The current density in the L-region as a function of $\tilde{%
\protect\alpha}$ for Dirichlet, Neumann and Robin (with $\protect\beta %
/a=-1/2$) boundary conditions (left panel) and as a function of the
coefficient in Robin boundary condition (right panel). The graphs are
plotted for $z_{0}/L=1$, $z/z_{0}=0.8$. On the left panel the dashed curve
corresponds to the current density in the geometry without the brane. The
vertical dotted line on the right panel corresponds to the critical value $%
\protect\beta _{c}/a\approx 0.228$.}
\label{fig6}
\end{figure}

The left panel of figure \ref{fig6} displays the vacuum current density in
the L-region for Dirichlet, Neumann and Robin (with $\beta /a=-1/2$)
boundary conditions as a function of the phase in the quasiperiodicity
condition along the compact dimensions. Recall that in the numerical
evaluations we consider a $D=4$ minimally coupled massless field. In this
case $\nu =2$ and the modes with purely imaginary $\lambda $ and the special
mode with $\lambda =0$ are absent for $a/\beta <4$. The graphs are plotted
for $z_{0}/L=1$, $z/z_{0}=0.8$. The dashed curve corresponds to the current
density in the absence of the brane. The right panel of figure \ref{fig6}
presents the dependence of the ratio $n_{l}\langle j^{l}\rangle /\langle
j^{l}\rangle _{\mathrm{M}}$ on $\beta /a$. The horizontal dashed lines
correspond to Dirichlet and Neumann boundary conditions. The vertical dotted
line corresponds to the critical value $\beta _{c}/a\approx 0.228$. In the
region $0<\beta <\beta _{c}$ the vacuum is unstable.

\begin{figure}[tbph]
\begin{center}
\begin{tabular}{cc}
\epsfig{figure=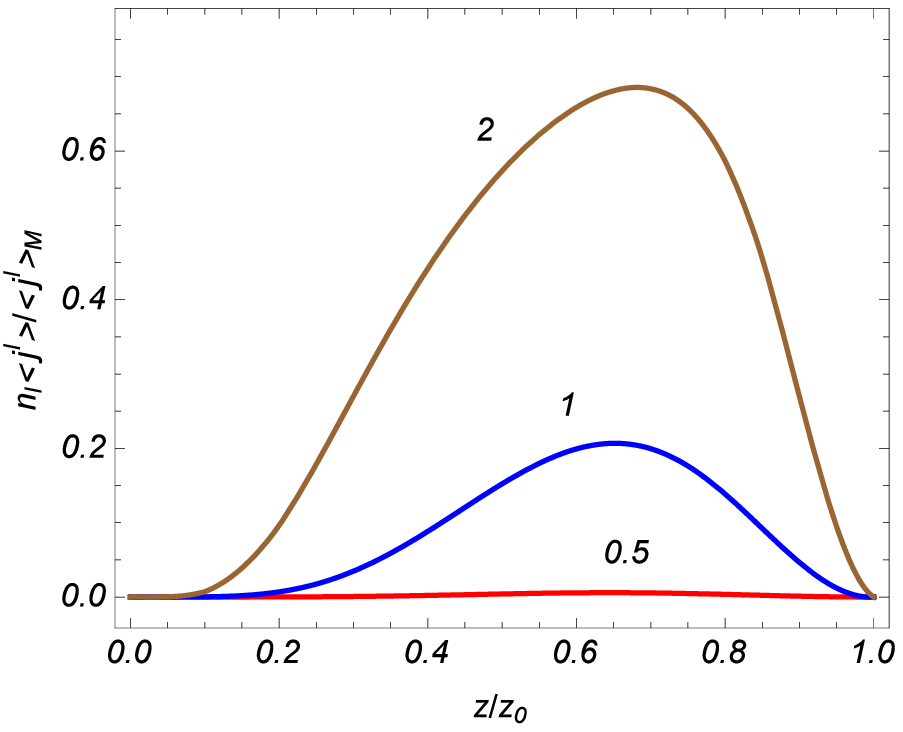,width=7.cm,height=5.5cm} & \quad %
\epsfig{figure=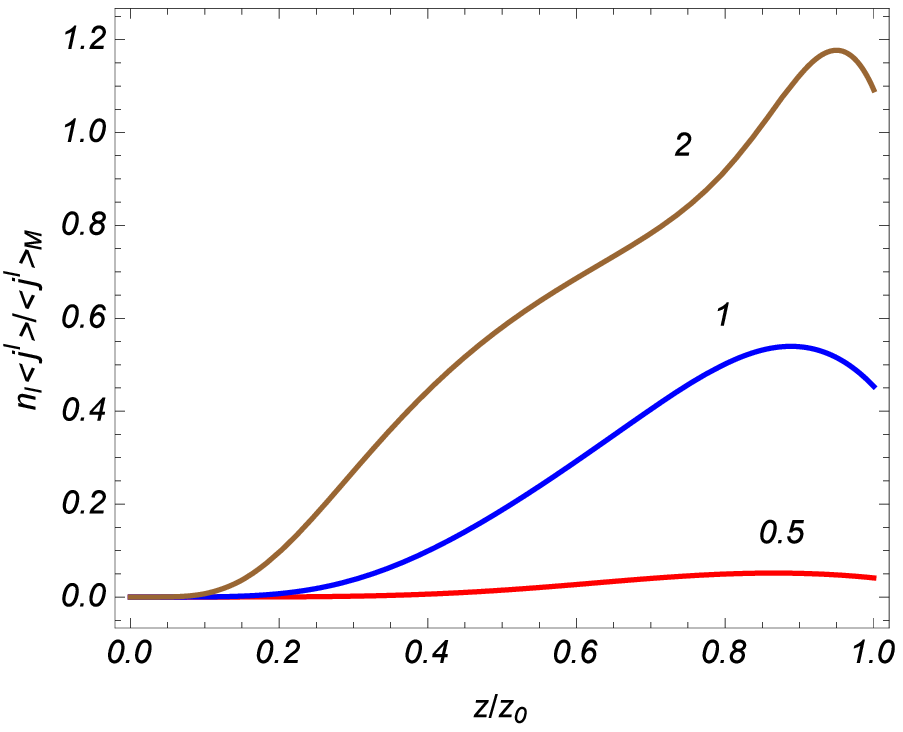,width=7.cm,height=5.5cm}%
\end{tabular}%
\end{center}
\caption{The ratio $n_{l}\langle j^{l}\rangle /\langle j^{l}\rangle _{%
\mathrm{M}}$ versus $z/z_{0}$ for Dirichlet (left panel) and Neumann (right
panel) boundary conditions. The figures near the curves correspond to the
values of $z_{0}/L$ and for the phase we have taken $\tilde{\protect\alpha}=%
\protect\pi /2$.}
\label{fig7}
\end{figure}

In figure \ref{fig7}, for $\tilde{\alpha}=\pi /2$, we show the dependence of
the ratio $n_{l}\langle j^{l}\rangle /\langle j^{l}\rangle _{\mathrm{M}}$ on
$z/z_{0}$ for Dirichlet (left panel) and Neumann (right panel) boundary
conditions for separate values of $z_{0}/L$ (numbers near the curves). For
Dirichlet condition the current density vanishes on the AdS boundary and on
the brane. As it follows from the asymptotic (\ref{jlbAdSbound}), near the
AdS boundary the charge flux density $n_{l}\langle j^{l}\rangle $ behaves as
$z^{D+2\nu +1}$. For the Minkowskian VEV with the length of the compact
dimension $aL/z$, equal to the proper length on the AdS bulk, one has $%
\langle j^{l}\rangle _{\mathrm{M}}\propto z^{D+1}$. Hence, the ratio plotted
in figure \ref{fig7} vanishes on the AdS boundary as $z^{2\nu }$. Similar
graphs in the case of Robin boundary condition are presented in figure \ref%
{fig8} for several values of $\beta /a$ (numbers near the curves). Note that
for all the examples in the L-region there are no modes with purely
imaginary $\lambda $.

\begin{figure}[tbph]
\begin{center}
\epsfig{figure=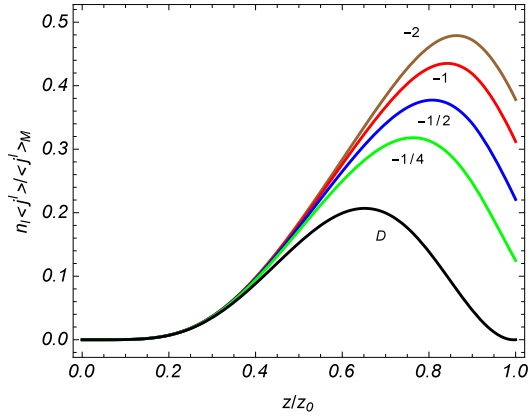,width=7.cm,height=5.5cm}
\end{center}
\caption{The same as in figure \protect\ref{fig6} for the fixed value $%
z/z_{0}=1$ in the case of Robin boundary condition. The numbers near the
curves correspond to the values of $\protect\beta /a$.}
\label{fig8}
\end{figure}

\section{Conclusion}

\label{sec:Conc}

We have studied the effects induced by a brane, parallel to the AdS
boundary, on the VEV of the current density for a massive charged scalar
field with an arbitrary curvature coupling parameter. The background
geometry under consideration is obtained from the $(D+1)$-dimensional AdS
one, described in Poincar\'{e} coordinates, by a toroidal compactification
of a part of spatial dimensions. Along compact dimensions the field operator
obeys quasiperiodicity conditions with general constant phases. We have also
assumed the presence of a constant gauge field. By the gauge transformation
the problem is reduced to the one with a zero gauge field. In the new gauge
the phases in the periodicity conditions are shifted by an amount determined
by the ratio of the magnetic flux, enclosed by a compact dimension, to the
flux quantum. On the brane and on the AdS boundary the field operator is
constrained by Robin and Dirichlet boundary conditions, respectively. We
consider a non-interacting quantum field and the Hadamard two-point function
contains all the information about the properties of the vacuum state.
Though the background geometry is homogeneous, the extrinsic curvature
tensor of the brane is nonzero and, as a consequence of this, the regions on
the right (R-region) and on the left (L-region) of the brane are not
physically equivalent.

In the R-region the spectrum of the quantum number $\lambda $ is continuous
and the mode functions obeying the boundary condition on the brane have the
form (\ref{Modes}) with the radial function (\ref{Z}). In addition to these
modes, under the condition $a/\beta >\nu -D/2$, one can have a bound state
with the mode function (\ref{BS}) and with the eigenvalue for $\eta $ being
the root of the equation (\ref{BSval}). The root should be constrained by $%
\eta <k_{(q)}^{(0)}$, where $k_{(q)}^{(0)}$ is defined as (\ref{kq0}). For
the modes with $\eta >k_{(q)}^{(0)}$ the energy becomes imaginary and they
lead to the instability of the Poincar\'{e} vacuum state. In addition to the
parameters $\nu $ and $k_{(q)}^{(0)}$, the stability of the vacuum depends
on the location of the brane. Depending on the value of the Robin
coefficient, the approaching of the brane to AdS boundary can lead to the
instability. For the value of the Robin coefficient given by (\ref{betsp})
there is a special mode (\ref{ModeSp}) for which $\lambda =0$. For a
minimally coupled massless scalar field the special value of the Robin
coefficient corresponds to Neumann boundary condition and the mode function
does not depend on the radial coordinate $z$. This is an analog of the
graviton zero mode in Randall--Sundrum 1-brane model. The Hadamard function
in the R-region is given by the expression (\ref{GL3}) with the second term
in the right-hand side being the brane-induced contribution. We have shown
that this expression holds in the presence of bound states as well.
Alternative representations for the Hadamard function in the R-region, (\ref%
{GLalt}) and (\ref{Gl}), are derived in Appendix, by using the summation
formula (\ref{AP2}).

In the L-region the spectrum for $\lambda $ is discrete and its eigenvalues
are roots of the equation (\ref{Jzer}). The corresponding expression for the
Hadamard function contains series over these roots. Another expression, in
which the explicit knowledge of the eigenvalues for $\lambda $ is not
required is obtained by making use of the generalized Abel--Plana formula (%
\ref{sumAP}). This allows us to extract manifestly the brane-induced
contribution, given by the second term in the right-hand side of (\ref{GR2}%
). For $a/\beta >D/2+\nu $ there is a mode with purely imaginary $\lambda
=i\eta $ which is the root of the equation (\ref{Immodes}) with $x=\eta
z_{0} $. For the stability of the vacuum state one needs to have $\eta
<k_{(q)}^{(0)}$. The stability condition depends also on the position of the
brane. The shift of the brane to the direction of the AdS boundary gives
rise to the instability in the initially stable vacuum state. An alternative
expression for the Hadamard function in the L-region is given by (\ref{GRalt}%
).

In both the R- and L-regions the VEV\ of the current density is decomposed
into the boundary-free and brane-induced contributions. For both these
contributions, the component of the current density along the $l$th compact
dimension is an odd periodic function of the phase $\tilde{\alpha}_{l}$ and
an even periodic function of the phases $\tilde{\alpha}_{i}$, $i\neq l$,
with the period equal to $2\pi $. In the R-region the brane-induced
contribution is given by (\ref{jla}). We have checked that for large values
of the AdS curvature radius the leading term in the corresponding asymptotic
expansion coincides with the boundary-induced part of the current density in
Minkowski spacetime with topology $R^{p+1}\times T^{q}$ in the presence of a
single Robin plate. In the case of a conformally coupled massless field the
current density coincides, up to the conformal factor $(z/a)^{D+1}$, with
the corresponding quantity in Minkowski bulk with the plate at $z=z_{0}$ on
which the field obeys the Robin boundary condition with the coefficient $%
\beta _{\mathrm{M}}^{(+)}$ given by (\ref{betM}). For points near the AdS
horizon ($z$ is large compared with the other length scales), the
brane-induced contribution is suppressed by the factor $e^{-2zk_{(q)}^{(0)}}$%
. In the same limit the boundary-free part behaves as $(z/a)^{D+1}$ and it
dominates in the total VEV. In the limit when the brane approaches to the
AdS boundary, for fixed values of $z$ and $L_{i}$, the brane-induced
contribution tends to zero like $z_{0}^{2\nu }$. For the investigation of
the near brane asymptotic of the vacuum current it is more convenient to use
the representation (\ref{jlLalt}). In the presence of a bound state its
contribution should be added separately to the right-hand side. For Robin
boundary condition with (\ref{betsp}) an additional contribution, given by (%
\ref{jls}), comes from the special mode (\ref{ModeSp}). An important
conclusion which follows from the representation (\ref{jlLalt}) is that the
current density is finite on the brane. This behavior is in clear contrast
with that for the VEVs of the field squared and of the energy-momentum
tensor which diverge on the brane. For Dirichlet boundary condition the
current density and its normal derivative vanish for points on the brane.
The asymptotic behavior of the current density along $l$th compact dimension
for large values of the corresponding length $L_{l}$ crucially depends
whether the parameter $k_{(q-1)}^{(0)}$, defined by (\ref{kq-10}), is zero
or not. For $k_{(q-1)}^{(0)}\neq 0$ the current density decays exponentially
like $e^{-L_{l}k_{(q-1)}^{(0)}}$. In the case $k_{(q-1)}^{(0)}=0$ the decay
is power law, as $1/L_{l}^{p+2\nu +2}$, for both massless and massive
fields. This behavior for massive fields is essentially different from that
for Minkowski bulk where the current is suppressed exponentially, by the
factor $e^{-mL_{l}}$. The expression for the current density in the
generalized Randall--Sundrum 1-brane model with compact dimensions is
obtained from the formulas in section \ref{sec:Curr} with an additional
factor $1/2$ and with the Robin coefficient (\ref{betRS}) for untwisted
fields and with $\beta =0$ for twisted fields.

The current density in the L-region is given by the expression (\ref{jlb1}).
For a conformally coupled massless field, this expression is reduced to the
one in Minkowski spacetime with two parallel plates, multiplied by the
conformal factor $(z/a)^{D+1}$, with Dirichlet boundary condition on the
left plate and Robin condition on the right one. On the AdS boundary the
brane-induced contribution vanishes as $z^{D+2\nu +2}$. For a fixed
observation point, when the location of the brane tends to the AdS horizon,
the brane-induced effects are suppressed by the factor $%
e^{-2z_{0}k_{(q)}^{(0)}}$. A similar behavior is exhibited by the
boundary-free part. From an alternative representation (\ref{jlbalt}) it
follows that the current density on the brane is finite and vanishes for
Dirichlet boundary condition. For large values of the length $L_{l}$, unlike
the R-region, the decay of the current density in the L-region is
exponential for both cases $k_{(q-1)}^{(0)}=0$ and $k_{(q-1)}^{(0)}\neq 0$.
This feature is related to the discreteness of the spectrum for $\lambda $.
For the value (\ref{SpL}) of the Robin coefficient the contribution (\ref%
{jLs1}) from the special mode has to be added to the right-hand side of (\ref%
{jlbalt}). For large values of $L_{l}$, this contribution dominates in the
VEV of the total current density. In particular, in the model with a single
compact dimension its decay, as a function of $L_{l}$, is power law, as $%
1/L_{l}^{D-1}$. For the brane location near the AdS boundary, the asymptotic
of the current density is given by (\ref{jlBhor2}) and it is suppressed by
the factor $e^{-\gamma _{1}L_{l}/z_{0}}$. In the presence of the special
mode, for a given value of $z/z_{0}$ (fixed distance from the brane), the
corresponding contribution to the current density behaves as $%
(z_{0}/L_{l})^{D-2}$ and, hence, it dominates in the total VEV.

The numerical results have been presented for the $D=4$ model with a single
compact dimension and for a minimally coupled massless scalar field. These
results show that, depending on the value of the Robin coefficient, the
presence of the brane can either increase or decrease the current density.
In particular, in the example considered, the modulus of the current density
takes its minimal value for Dirichlet boundary condition.

\section*{Acknowledgments}

This work has been supported in part by the State Committee of Science of
the Ministry of Education and Science of the Republic of Armenia. A. A. S.
gratefully acknowledges the hospitality of the INFN, Laboratori Nazionali di
Frascati (Frascati, Italy), where part of this work was done.

\appendix

\section{Other representations of the two-point function}

Here we provide representations for the Hadamard function convenient in the
investigation of near brane asymptoics for the VEV of the current density.
First we consider the R-region. In the representation (\ref{GL1}) we
separate the series over $n_{l}$ and for the summation use the Abel--Plana
type formula \cite{Bell10,Beze08}%
\begin{eqnarray}
&&\frac{2\pi }{L_{l}}\sum_{n_{l}=-\infty }^{\infty
}g(k_{l})f(|k_{l}|)=\int_{0}^{\infty }du[g(u)+g(-u)]f(u)  \notag \\
&&\qquad +i\int_{0}^{\infty }du\,[f(iu)-f(-iu)]\sum_{s=\pm 1}\frac{g(isu)}{%
e^{uL_{l}+is\tilde{\alpha}_{l}}-1},  \label{AP2}
\end{eqnarray}%
with $k_{l}$ defined in (\ref{kl}) (formula (\ref{AP2}) is reduced to the
standard Abel--Plana formula in the special case $g(x)=1$, $\tilde{\alpha}%
_{l}=0$). After the application (\ref{AP2}) with $g(k_{l})=e^{ik_{l}\Delta
x^{l}}$, the Hadamard function is presented in the form%
\begin{equation}
G(x,x^{\prime })=G_{R^{p+2}\times T^{q-1}}(x,x^{\prime })+G_{l}(x,x^{\prime
}).  \label{Gdecl}
\end{equation}%
Here the part $G_{R^{p+2}\times T^{q-1}}(x,x^{\prime })$ comes from the
first term in the right-hand side of (\ref{AP2}) and is the Hadamard
function for the geometry with a single brane in $(D+1)$-dimensional AdS
spacetime with spatial topology $R^{p+2}\times T^{q-1}$ for which the $l$th
dimension is decompactified. The lengths of the remaining compact dimensions
are the same: $(L_{p+1},\ldots ,L_{l-1},L_{l+1},\ldots ,L_{D-1})$. The
second term on the right of (\ref{Gdecl}) is induced by the compactification
of the $l$th dimension and is given by the expression%
\begin{eqnarray}
G_{l}(x,x^{\prime }) &=&\frac{4\left( zz^{\prime }\right) ^{D/2}L_{l}}{%
\left( 2\pi \right) ^{p+1}a^{D-1}V_{q}}\sum_{n=1}^{\infty }\sum_{\mathbf{n}%
_{q-1}}\int d\mathbf{k}_{p}\,e^{ik_{r}\Delta x^{r}}\int_{0}^{\infty
}d\lambda \,\lambda  \notag \\
&&\times \frac{g_{\nu }(\lambda z_{0},\lambda z)g_{\nu }(\lambda
z_{0},\lambda z^{\prime })}{\bar{J}_{\nu }^{2}(\lambda z_{0})+\bar{Y}_{\nu
}^{2}(\lambda z_{0})}\,\int_{0}^{\infty }dw\cosh (w\Delta t)  \notag \\
&&\times \frac{e^{-nuL_{l}}}{u}\cosh (u\Delta x^{l}+in\tilde{\alpha}%
_{l})|_{u=\sqrt{w^{2}+\lambda ^{2}+k^{(l)2}}},  \label{GLalt}
\end{eqnarray}%
where $\mathbf{n}_{q-1}=(n_{p+1},\ldots ,n_{l-1},n_{l+1},\ldots ,n_{D-1})$, $%
k^{(l)2}=k^{2}-k_{l}^{2}$, and the summation over $r$ in the exponent goes
over $r=1,\ldots ,D-1$, $r\neq l$. In deriving this result, we have used the
relation%
\begin{equation}
\sum_{s=\pm 1}\frac{e^{-su\Delta x^{l}}}{e^{uL_{l}+is\tilde{\alpha}_{l}}-1}%
=2\sum_{n=1}^{\infty }e^{-nuL_{l}}\cosh (u\Delta x^{l}+in\tilde{\alpha}_{l}).
\label{Rel1}
\end{equation}%
Note that the part $G_{R^{p+1}\times T^{q-1}}(x,x^{\prime })$ does not
contribute to the VEV of the current density along $l$th dimension. The
expression (\ref{GLalt}) gives the contribution to the Hadamard function
from the modes with continuous $\lambda $. In the presence of a bound state,
a similar representation can be found for the corresponding contribution.

Another useful representation for the Hadamard function in the R-region is
obtained by using the identity (\ref{Ident}) for the integrand in (\ref%
{GLalt}). The part in $G_{l}(x,x^{\prime })$ coming from the first term in
the right-hand side of (\ref{Ident}) gives the corresponding function in the
boundary-free AdS spacetime (the brane is absent), denoted here by $%
G_{0l}(x,x^{\prime })$. In the part induced by the brane, coming from the
last term in (\ref{Ident}), we rotate the integration contour by the angle $%
\pi /2$ for the term with $j=1$ and by the angle $-\pi /2$ for the term with
$j=2$. In this way we get%
\begin{eqnarray}
G_{l}(x,x^{\prime }) &=&G_{0l}(x,x^{\prime })-\frac{a^{1-D}L_{l}(zz^{\prime
})^{D/2}}{2^{p-1}\pi ^{p+2}V_{q}}\sum_{n=1}^{\infty }\sum_{\mathbf{n}%
_{q-1}}\int d\mathbf{k}_{p}\,e^{ik_{r}\Delta x^{r}}  \notag \\
&&\times \int_{0}^{\infty }dw\,\cosh (w\Delta t)\int_{0}^{\infty
}du\,\sum_{s=\pm 1}\cos (u\Delta x^{l}-sn\tilde{\alpha}_{l})e^{isnuL_{l}}
\notag \\
&&\times \frac{\bar{I}_{\nu }(\lambda z_{0})}{\bar{K}_{\nu }(\lambda z_{0})}%
K_{\nu }(\lambda z)K_{\nu }(\lambda z^{\prime })|_{\lambda =\sqrt{%
w^{2}+u^{2}+k^{(l)2}}}.  \label{Gl}
\end{eqnarray}

Now, we consider the L-region. The corresponding expression for the Hadamard
function is given by (\ref{GR}). In a way similar to that for the R-region,
applying the formula (\ref{AP2}), for the part induced by the
compactification of the $l$th dimension we find the following representation%
\begin{eqnarray}
G_{l}(x,x^{\prime }) &=&\frac{8\left( zz^{\prime }\right) ^{D/2}L_{l}}{%
\left( 2\pi \right) ^{p+1}a^{D-1}V_{q}z_{0}^{2}}\sum_{n=1}^{\infty }\sum_{%
\mathbf{n}_{q-1}}\int d\mathbf{k}_{p}\,e^{ik_{r}\Delta
x^{r}}\sum_{n}\,\gamma _{n}  \notag \\
&&\times T_{\nu }(\gamma _{n})J_{\nu }(\gamma _{n}z/z_{0})J_{\nu }(\gamma
_{n}z^{\prime }/z_{0})\int_{0}^{\infty }dw\,\cosh (w\Delta t)  \notag \\
&&\times \frac{e^{-nuL_{l}}}{u}\cosh (u\Delta x^{l}+in\tilde{\alpha}%
_{l})|_{u=\sqrt{w^{2}+\gamma _{n}^{2}/z_{0}^{2}+k^{(l)2}}}.  \label{GRalt}
\end{eqnarray}%
In this expression, for the summation over $n$ we can use the formula (\ref%
{sumAP}). The part coming from the first term in the right-hand side of (\ref%
{sumAP}) corresponds to the boundary free contribution and, as a result, we
obtain the decomposition similar to (\ref{Gl}) where now the brane-induced
contribution is obtained from that in (\ref{Gl}) by the replacements $I_{\nu
}(x)\rightleftarrows $ $K_{\nu }(x)$.

\end{document}